%% file: main.tex
\pdfoutput=1
\documentclass[11pt,letterpaper]{article}

\usepackage[margin=1in]{geometry}
\usepackage[T1]{fontenc}
\usepackage[utf8]{inputenc}
\usepackage{lmodern}
\usepackage{microtype}
\usepackage{amsmath,amssymb,amsthm,mathtools,bm}
\usepackage{braket}
\usepackage{graphicx}
\usepackage{subcaption}
\usepackage{booktabs}
\usepackage{array}
\usepackage{tabularx}
\usepackage{enumitem}
\usepackage{xcolor}
\usepackage{float}
\usepackage[section]{placeins}
\usepackage{tikz}
\usetikzlibrary{arrows.meta,positioning,fit,shapes.geometric,decorations.pathreplacing,calc}
\usepackage[numbers,sort&compress]{natbib}
\usepackage{hyperref}
\usepackage[nameinlink,capitalize,noabbrev]{cleveref}
\usepackage{fancyhdr}
\usepackage{titlesec}
\usepackage{setspace}
\usepackage{caption}
\newcolumntype{P}[1]{>{\raggedright\arraybackslash}p{#1}}
\newcolumntype{Z}{>{\raggedright\arraybackslash}X}

\definecolor{uwblue}{RGB}{0,74,136}
\definecolor{softgray}{RGB}{246,247,249}
\definecolor{softblue}{RGB}{238,245,252}
\hypersetup{
  colorlinks=true,
  linkcolor=uwblue,
  citecolor=uwblue,
  urlcolor=uwblue,
  pdftitle={Conditional Information Flow and Neural Representability of One-Dimensional Rotated Cluster States},
  pdfauthor={Ziqi Xu}
}
\graphicspath{{figures/}}
\titleformat{\section}{\Large\bfseries\color{uwblue}}{\thesection}{0.7em}{}
\titleformat{\subsection}{\large\bfseries}{\thesubsection}{0.7em}{}
\titleformat{\subsubsection}{\normalsize\bfseries}{\thesubsubsection}{0.7em}{}
\titlespacing*{\section}{0pt}{2.2ex plus 0.5ex minus 0.2ex}{1.0ex}
\titlespacing*{\subsection}{0pt}{1.8ex plus 0.4ex minus 0.2ex}{0.7ex}

\newcommand{\E}{\mathbb{E}}
\newcommand{\Tr}{\operatorname{Tr}}
\newcommand{\rank}{\operatorname{rank}}
\newcommand{\spec}{\operatorname{spec}}
\newcommand{\xitwo}{\xi_{2\mathrm{pt}}}
\newcommand{\xicmi}{\xi_{\mathrm{CMI}}}
\newcommand{\id}{\mathbb{I}}

\begin{document}

\begin{center}
{\Large\bfseries
Conditional Information Flow and Neural Representability\\[0.2em]
of One-Dimensional Rotated Cluster States\par}
\vspace{1.2em}
{\large Ziqi Xu\par}
\vspace{0.6em}
{\small\itshape
Department of Physics and Astronomy, University of Waterloo, Waterloo, Ontario N2L 3G1, Canada\\
Perimeter Institute for Theoretical Physics, Waterloo, Ontario N2L 2Y5, Canada\par}
\vspace{0.4em}
{\small October 2026\par}
\end{center}

\vspace{0.6em}
\begin{abstract}
\noindent
I study how measurement outcomes can remain conditionally dependent over long distances while the underlying state carries only a small internal memory, using one-dimensional rotated cluster states, whose exact matrix product state (MPS) bond dimension is $D=2$ at every rotation angle. Using exact MPS contraction, I compute the conditional mutual information (CMI) of two readout protocols: the entanglement-swapping endpoint CMI decreases from one bit at $\theta=0$ to zero at $\theta=\pi/2$, whereas a stationary bulk readout of the locally equivalent conventional cluster state vanishes at both limits. As an exact control, the spin-1 AKLT chain obeys $I_m=(4/9)\,3^{-m}$, so its CMI length equals its two-point correlation length, for an identifiable but non-generic reason. I then compare learners on the same $N=64$, $\theta=0.35$ target. A positive autoregressive GRU learns the computational-basis Born distribution from a few thousand samples, but an exactly computable Bell-basis readout shows that this distribution does not fix the phases. A complex RBM reaches the $N=8$ ground state under an exact objective, while a short generic recurrent variational Monte Carlo pilot is still improving when its runs end. As a positive control, an MPS-induced recurrent model whose hidden state is the bond vector reaches the ground state to near machine precision at $\chi=2$, whereas a $\chi=1$ control fails. These results show that long CMI does not require a large bond dimension, and that building the small virtual memory into the architecture greatly eases optimization.

\medskip
\noindent\textit{This manuscript is the author's M.Sc.\ research paper, University of Waterloo (2026).}
\end{abstract}

\vspace{0.8em}
\thispagestyle{plain}

\input{sections/01_introduction}
\input{sections/02_framework}
\input{sections/03_mps}

\input{sections/04_rbm_rnn_results}
\input{sections/05_geometry_and_implications}
\input{sections/07_discussion_conclusion}

\section*{Code availability}
\addcontentsline{toc}{section}{Code availability}
The code used to generate the data and figures in this paper is available from the author upon reasonable request.

\section*{Acknowledgements}
\addcontentsline{toc}{section}{Acknowledgements}
I thank my supervisor, Robert Mann, for his guidance and for the freedom to explore different directions during this project, and Tim Hsieh for reading the manuscript and for his comments and suggestions. I am also grateful to the University of Waterloo and the Perimeter Institute for Theoretical Physics for their research environment.

\appendix
\input{sections/06_methods_and_limitations}

\bibliographystyle{unsrtnat}
\bibliography{references}

\end{document}

%% file: sections/01_introduction.tex
\section{Introduction}

\subsection{Question and motivation}

A quantum state can have a small exact representation and still produce measurement outcomes with a long memory. The bond dimension of an exact matrix product state (MPS) counts how many virtual degrees of freedom are needed at one cut. Conditional mutual information (CMI) measures how long some of that information remains useful after a sequence of measurements. The two quantities are different: a single bit can be carried over many sites.

I study this distinction using rotated one-dimensional cluster states. One-dimensional cluster states are stabilizer states with exact bond-dimension-two MPS representations
\cite{RaussendorfBriegel2001,RaussendorfBrowneBriegel2003,Hein2006,Schollwock2011}. They are therefore simple enough for exact calculations, but their measured distributions are not always short-memory processes. This makes them a useful test for neural quantum states: the target state is compact, yet a generic model still has to discover the correct internal variable and phase rule.

A general $N$-qubit pure state is
\begin{equation}
    \ket{\psi}=\sum_{x\in\{0,1\}^{N}}\psi(x)\ket{x},
    \qquad
    \sum_x |\psi(x)|^2=1.
    \label{eq:state-expansion}
\end{equation}
Here $x=(x_1,\ldots,x_N)$ is a binary measurement string and $\psi(x)$ is its complex amplitude. The Born rule gives the probability of observing that string in the computational basis,
\begin{equation}
    p(x)=|\psi(x)|^2.
    \label{eq:born-rule}
\end{equation}
A direct table of all amplitudes contains $2^N$ complex numbers. The practical problem is to replace that table by a smaller structured model. In this paper the models are a restricted Boltzmann machine (RBM), recurrent neural networks (RNNs), and MPS.

\subsection{Relation to previous work and contributions}
\label{sec:prior-work}

The closest previous work is Ref.~\cite{Yang2024}, which argued that the conditional correlations of a measurement distribution, rather than its ordinary few-body correlations, control how efficiently a neural network can represent a quantum state. That work introduced the rotated entanglement-swapping cluster chain studied here, proved the upper bound on its endpoint CMI quoted in \cref{eq:yang-bound}, noted that the bond-dimension-two MPS of this state maps onto a recurrent network, and reported that variational RNN and RBM states at $N=64$ become harder to optimize as the rotation angle approaches zero. Recurrent architectures built directly from tensor-network contractions were studied in Ref.~\cite{Wu2023}. The present paper builds on these results.

Conditional mutual information is also a central diagnostic in other settings. Its decay length, often called the Markov length, controls approximate quantum Markov chains~\cite{FawziRenner2015} and characterizes the stability of mixed-state phases~\cite{SangHsieh2025}. Entanglement generated by measuring part of a state can probe its sign structure~\cite{Lin2023SignMIE} and can become long ranged in shallow random circuits~\cite{BaoBlockAltman2024}. A closely related older notion is the localizable entanglement, the entanglement that local measurements can concentrate between two distant sites~\cite{VerstraetePoppCirac2004,Popp2005}; its length can exceed the ordinary correlation length, with the AKLT chain as a standard example connected to string order~\cite{denNijsRommelse1989,KennedyTasaki1992}. The CMI of a measured chain studied here is a classical counterpart of these quantities: it asks how much information about one endpoint survives in the measurement record rather than in a post-measurement quantum state.

Beyond Ref.~\cite{Yang2024}, this paper computes the CMI exactly over the full rotation range for two readout protocols, a finite-endpoint entanglement-swapping readout and a stationary bulk readout of the conventional cluster state, and compares finite-window CMI lengths with the length scale of the known bound (\cref{sec:cmi-scan}). It derives the exact CMI of the spin-1 AKLT chain, $I_m=(4/9)\,3^{-m}$, and explains why its CMI length coincides with its two-point correlation length (\cref{sec:aklt-control}). It describes the angle dependence in terms of the singular values of the measurement maps on the virtual bond, and uses an $N=8$ Bell-basis readout to show explicitly that computational-basis statistics do not fix the phases (\cref{sec:angle-interpretation}). Finally, it compares four learners on the same target: a complex RBM with an exact objective, a GRU trained only on measurement strings, a generic recurrent variational Monte Carlo (VMC) ansatz, and an MPS-induced recurrent model used as a positive control together with a $\chi=1$ negative control (\cref{sec:mps-induced-model,sec:neural-results}).

\subsection{Physical model}

The single-qubit Pauli matrices are
\begin{equation}
 X=\begin{pmatrix}0&1\\1&0\end{pmatrix},\qquad
 Y=\begin{pmatrix}0&-i\\i&0\end{pmatrix},\qquad
 Z=\begin{pmatrix}1&0\\0&-1\end{pmatrix}.
 \label{eq:pauli-matrices}
\end{equation}
A subscript indicates the site on which a matrix acts; identities on all other sites are left implicit. Throughout the paper, $\id$ denotes the identity matrix, $\dagger$ denotes complex-conjugate transpose, $\Tr$ denotes a matrix trace, and $\operatorname{Re}$ denotes the real part of a complex number. Logarithms in entropies are base two; logarithms used to define decay lengths are natural logarithms. For an open one-dimensional cluster chain with an even number of sites, I use two locally equivalent cluster Hamiltonians. For a graph state, the stabilizer generator associated with vertex \(v\) is
\begin{equation}
K_v = X_v \prod_{u\in N(v)} Z_u,
\end{equation}
where \(N(v)\) denotes the set of vertices neighboring \(v\). Thus the conventional Hamiltonian form is
\begin{equation}
 H_{\mathrm C}
 =-X_1Z_2
 -\sum_{j=2}^{N-1}Z_{j-1}X_jZ_{j+1}
 -Z_{N-1}X_N,
 \label{eq:HC-intro}
\end{equation}
and the entanglement-swapping form is
\begin{equation}
 \begin{split}
 H_{\mathrm{ES}}
 ={}&-Z_1X_2
 -\sum_{j=2}^{N-2}X_{j-1}Z_jX_{j+1}\\
 &-X_{N-1}X_N-X_{N-2}Z_{N-1}Z_N.
 \end{split}
 \label{eq:HES-intro}
\end{equation}
Each Hamiltonian is a sum $H=-\sum_{j=1}^{N}K_j$ of $N$ mutually commuting stabilizers. Every stabilizer satisfies $K_j^2=\id$, so its eigenvalues are $\pm1$. The unique open-chain ground state satisfies $K_j\ket{\psi_0}=\ket{\psi_0}$ for every $j$, and its exact energy is therefore
\begin{equation}
 E_0=-N.
 \label{eq:exact-energy-intro}
\end{equation}
This simple formula is used repeatedly as an optimization check.

The two Hamiltonians are related by local Hadamard gates on the first $N-1$ sites. Since a Hadamard exchanges $X$ and $Z$, the spectra are identical, but computational-basis measurements are not identical: the local unitary also changes which physical basis is being read out. This is why I keep the two measurement protocols separate.

I rotate every site around the $y$ axis using
\begin{equation}
 R_y(\theta)=e^{-i\theta Y/2}
 =\begin{pmatrix}
 \cos(\theta/2)&-\sin(\theta/2)\\
 \sin(\theta/2)& \cos(\theta/2)
 \end{pmatrix}.
 \label{eq:Ry-intro}
\end{equation}
The rotated Hamiltonian and ground state are
\begin{equation}
 H(\theta)=R_y(\theta)^{\dagger\otimes N}H(0)R_y(\theta)^{\otimes N},
 \qquad
 \ket{\psi_\theta}=R_y(\theta)^{\dagger\otimes N}\ket{\psi_0}.
 \label{eq:rotated-H-state}
\end{equation}
Unitary conjugation leaves the spectrum unchanged, so the ground-state energy is still $-N$. Because the rotation is local, it also leaves the Schmidt rank across every cut unchanged. The exact cluster-state bond dimension therefore remains $D=2$ for every $N$ and $\theta$.

\subsection{Conditional mutual information and length scales}
\label{sec:cmi-lengths}

For a discrete random variable $A$ with distribution $p(a)$, the Shannon entropy in bits is
\begin{equation}
 H(A)=-\sum_a p(a)\log_2 p(a).
 \label{eq:shannon-entropy}
\end{equation}
It measures uncertainty before the outcome is known. The mutual information between $A$ and $C$ is
\begin{equation}
 I(A:C)=H(A)+H(C)-H(A,C),
 \label{eq:mutual-information}
\end{equation}
which is the reduction in uncertainty about one variable after the other is known. The conditional mutual information is
\begin{equation}
 I(A:C\mid B)
 =H(A,B)+H(B,C)-H(B)-H(A,B,C).
 \label{eq:cmi-entropy}
\end{equation}
Equivalently,
\begin{equation}
 I(A:C\mid B)
 =\sum_{a,b,c}p(a,b,c)
 \log_2\!\frac{p(a,b,c)p(b)}{p(a,b)p(b,c)}.
 \label{eq:cmi-probability}
\end{equation}
The two formulas are the same; the second one is convenient for numerical evaluation from a joint probability table \cite{CoverThomas2006}. CMI also has a direct predictive meaning,
\begin{equation}
 I(A:C\mid B)
 =\sum_{a,b}p(a,b)\,
 D_{\mathrm{KL}}\!\left[p(C\mid a,b)\,\Vert\,p(C\mid b)\right].
 \label{eq:cmi-predictive}
\end{equation}
Here $D_{\mathrm{KL}}(p\Vert q)=\sum_c p(c)\log_2[p(c)/q(c)]$. Thus CMI is the average change in the prediction of $C$ when $A$ is added after $B$ is already known. This is the interpretation used throughout the measured-chain analysis.

For a measured chain, I place $A$ and $C$ at the two ends of a buffer of length $m$ and define
\begin{equation}
 I_m=I(X_0:X_{m+1}\mid X_1,\ldots,X_m).
 \label{eq:cmi-sequence}
\end{equation}
After the middle outcomes are known, $I_m$ tells us how much extra information the left endpoint gives about the right endpoint. If a clean tail follows
\begin{equation}
 I_m\simeq A e^{-m/\xicmi},
 \label{eq:cmi-exponential}
\end{equation}
where $A>0$ is the overall signal amplitude, then $\xicmi$ is the CMI length. Taking a logarithm gives
\begin{equation}
 \log I_m\simeq \log A-\frac{m}{\xicmi},
 \label{eq:cmi-log-fit}
\end{equation}
so $\xicmi$ is the negative inverse slope of $\log I_m$ against $m$. A useful local check is the effective length
\begin{equation}
 \xi_{\mathrm{eff}}(m)
 =-\frac{\Delta m}{\ln[I_{m+\Delta m}/I_m]}.
 \label{eq:cmi-effective-length}
\end{equation}
I use $\Delta m=1$ for a one-site sequence and $\Delta m=2$ when only one parity of buffer length is compared. If the decay is really exponential, $\xi_{\mathrm{eff}}(m)$ should approach a plateau. A constant nonzero $I_m$ corresponds to an infinite length. If $I_m$ is zero, or already at numerical precision, no decay length should be inferred from it.

The ordinary connected correlation function of a local observable $O$ is
\begin{equation}
 C_O(r)=\langle O_jO_{j+r}\rangle
       -\langle O_j\rangle\langle O_{j+r}\rangle.
 \label{eq:connected-correlation}
\end{equation}
When $|C_O(r)|\sim e^{-r/\xitwo}$, the parameter $\xitwo$ is the usual two-point correlation length. There is no general reason for $\xitwo$ and $\xicmi$ to agree. They average and condition the measurement record in different ways. The AKLT calculation in \cref{sec:aklt-control} gives one special case where they are equal and explains why.

It is also useful to distinguish the decay of ordinary correlations from the decay of conditional correlations. To state this more generally, consider a tripartite quantum state $\rho_{ABC}$, with von Neumann entropy
\begin{equation}
S(A)=-\operatorname{Tr}\rho_A\log_2\rho_A .
\end{equation}
The quantum mutual information and conditional mutual information are
\begin{equation}
I(A:C)=S(A)+S(C)-S(AC),
\end{equation}
and
\begin{equation}
I(A:C|B)
=
S(AB)+S(BC)-S(B)-S(ABC).
\end{equation}
For a classical measurement distribution, these expressions reduce to the Shannon-entropy formulas above when the density matrix is diagonal in the measurement basis.

For a pure tripartite state $\rho_{ABC}=|\psi\rangle\langle\psi|$, the Schmidt decomposition implies
\begin{equation}
S(AB)=S(C),\qquad
S(BC)=S(A),\qquad
S(AC)=S(B),
\end{equation}
together with $S(ABC)=0$. Therefore,
\begin{align}
I(A:C|B)
&=S(C)+S(A)-S(B) \\
&=S(A)+S(C)-S(AC) \\
&=I(A:C).
\end{align}
Thus, when $ABC$ is globally pure, the two quantities have identical dependence on the separation between $A$ and $C$: decay of the mutual information necessarily gives the same decay of the CMI.

This conclusion does not hold for a mixed state. Defining the tripartite information
\begin{equation}
\begin{aligned}
I_3(A:B:C)
={}&S(A)+S(B)+S(C) \\
&-S(AB)-S(AC)-S(BC)
+S(ABC),
\end{aligned}
\end{equation}
direct subtraction gives
\begin{equation}
I(A:C|B)=I(A:C)-I_3(A:B:C).
\end{equation}
For a pure tripartite state, the entropy identities above force $I_3=0$, whereas for a mixed state there is no such cancellation. In particular, $I_3$ is not fixed by the marginal mutual information $I(A:C)$. Consequently, knowing that $I(A:C)$ decreases with the separation between $A$ and $C$, even exponentially, is not sufficient in general to conclude that $I(A:C|B)$ decreases with the same length scale, or even that it decays at all. Conditioning on the intermediate region can therefore retain information that is not visible in ordinary marginal correlations.

\subsection{Organization of the numerical study}

The calculations are organized around three questions.

First, how does the measured CMI depend on the rotation angle and on the chosen readout? I answer this with exact MPS probabilities for both cluster protocols and with an AKLT control.

Second, how easily do different neural models learn the wavefunction or its Born distribution? The RBM and variational recurrent models minimize energy and therefore attempt to learn a complex state. The data-driven GRU only learns $p(x)$ from measurement strings. These are related but different tasks.

Third, does architecture matching explain the difference in optimization? To test this, I compare a generic recurrent ansatz with an MPS-induced recurrent model whose hidden state is already the exact type of two-dimensional virtual bond.

\begin{table}[!htbp]
\centering
\small
\caption{Numerical calculations used in this paper. The rows answer different questions and are not treated as one speed leaderboard.}
\label{tab:study-design}
\begin{tabularx}{\textwidth}{@{}P{0.20\textwidth}P{0.22\textwidth}P{0.15\textwidth}Z@{}}
\toprule
Calculation & Training or evaluation signal & Main size & Purpose \\
\midrule
Exact MPS CMI & Exact joint probabilities & finite and stationary chains & Measure conditional information, check angle limits, and separate bulk from endpoint readouts. \\
Complex RBM & Exact variational energy & $N=8$ & Test whether a generic global neural ansatz can represent the small state without Monte Carlo noise. \\
Data-driven GRU & Measurement strings & $N=64$ & Test how much data a small positive autoregressive model needs to learn the Born distribution. \\
Generic recurrent VMC & Sampled local energies & $N=64$ & Observe the convergence trend of a less structured complex recurrent ansatz. \\
MPS-induced recurrent model & Exact variational energy & $N=64$ & Test the advantage of building the known bond structure into the recurrent update. \\
AKLT control & Exact MPS probabilities & infinite-chain bulk & Give an analytic example with $\xicmi=\xitwo$. \\
\bottomrule
\end{tabularx}
\end{table}

The CMI fits are finite-window summaries of exact data, not asymptotic results, and the generic recurrent runs were stopped by a queue time limit before convergence, so they do not establish a final error floor.

%% file: sections/02_framework.tex
\section{RBM and recurrent neural networks}
\label{sec:neural-background}

This section introduces the RBM, a simple complex RNN for variational energy minimization, and a GRU for learning measurement probabilities. It also gives the variational-energy and sampling formulas used throughout the numerical experiments discussed below.

\subsection{Variational wavefunctions and local energy}

I write a model wavefunction as
\begin{equation}
 \psi_{\vartheta}(x)
 =\sqrt{p_{\vartheta}(x)}\,e^{i\phi_{\vartheta}(x)}.
 \label{eq:amplitude-phase}
\end{equation}
Here $\vartheta$ denotes all trainable parameters of the wavefunction. The magnitude determines the Born probability $p_{\vartheta}(x)$, while the real function $\phi_{\vartheta}(x)$ gives the phase. A data set collected in one orthonormal basis determines the probabilities in that basis, but not all relative phases \cite{Torlai2018,Carrasquilla2019}.

For a Hamiltonian $H$, the variational energy is the Rayleigh quotient
\begin{equation}
 E(\vartheta)
 =\frac{\bra{\psi_{\vartheta}}H\ket{\psi_{\vartheta}}}
        {\braket{\psi_{\vartheta}|\psi_{\vartheta}}}.
 \label{eq:variational-energy}
\end{equation}
The variational principle gives $E(\vartheta)\ge E_0$, where $E_0$ is the exact ground-state energy. Minimizing \cref{eq:variational-energy} is therefore a ground-state learning problem.

When the state is normalized, the same energy can be written as an average over configurations,
\begin{equation}
 E(\vartheta)=\sum_x p_{\vartheta}(x)E_{\mathrm{loc}}(x),
 \label{eq:energy-local-average}
\end{equation}
where
\begin{equation}
 E_{\mathrm{loc}}(x)
 =\sum_{x'}H_{x x'}
   \frac{\psi_{\vartheta}(x')}{\psi_{\vartheta}(x)},
 \qquad
 H_{x x'}=\bra{x}H\ket{x'}.
 \label{eq:local-energy}
\end{equation}
The local energy collects every basis state $x'$ connected to $x$ by the Hamiltonian. For a local spin Hamiltonian, only a small number of $x'$ contribute. Variational Monte Carlo (VMC) estimates \cref{eq:energy-local-average} from samples of $p_{\vartheta}(x)$. Because the expectation value is estimated from a finite set of sampled
configurations, the VMC energy has statistical noise even for fixed model
parameters.

The sampled energy gradient is often written with the logarithmic derivative
\begin{equation}
 O_k(x)=\frac{\partial}{\partial\vartheta_k}
        \ln\psi_{\vartheta}(x).
 \label{eq:log-derivative}
\end{equation}
For a real parameter coordinate $\vartheta_k$, the gradient is
\begin{equation}
 \frac{\partial E}{\partial\vartheta_k}
 =2\,\operatorname{Re}\!\left[
 \langle O_k^{*}E_{\mathrm{loc}}\rangle_p
 -\langle O_k^{*}\rangle_p\langle E_{\mathrm{loc}}\rangle_p
 \right].
 \label{eq:vmc-gradient}
\end{equation}
The brackets denote averages over $p_{\vartheta}$, and $\operatorname{Re}$ is the real part. In VMC, the expectation values in this covariance are estimated from a finite
Monte Carlo sample, so the resulting gradient and parameter updates inherit
sampling noise. For the full-basis RBM, the energy is computed by summing over all basis
states, while for the MPS it is obtained by exact tensor contraction. In
both cases, the objective is deterministic and its gradient is computed by
automatic differentiation.

There are two energy errors,
\begin{equation}
 \Delta E=E-E_0,
 \qquad
 \epsilon_E=\frac{|E-E_0|}{N}.
 \label{eq:energy-errors}
\end{equation}
The first is a total error and the second is an error per site.

\subsection{Restricted Boltzmann machine (RBM)}

An RBM contains visible spins $s_i\in\{-1,+1\}$ and hidden spins
$h_k\in\{-1,+1\}$. Here $x_i\in\{0,1\}$ denotes the computational-basis
bit on site $i$, and I encode the same physical configuration using
$s_i=1-2x_i$. I use $M$ hidden units for $N$ visible units. Before the hidden variables are summed out, an unnormalized complex amplitude can be written as
\begin{equation}
 \widetilde\psi_{\vartheta}(s)
 =\sum_{h\in\{-1,+1\}^{M}}
 \exp\!\left(
   \sum_{i=1}^{N}a_i s_i
  +\sum_{k=1}^{M}b_k h_k
  +\sum_{i=1}^{N}\sum_{k=1}^{M}W_{ik}s_i h_k
 \right).
 \label{eq:rbm-joint}
\end{equation}
The parameters $a_i$, $b_k$, and $W_{ik}$ are complex. Their real parts mainly control magnitudes and their imaginary parts can encode signs or phases. Because the hidden spins do not couple to one another, each hidden sum is independent. Summing over the hidden variables gives
\begin{equation}
 \log \widetilde\psi_{\vartheta}(s)
 =\sum_{i=1}^{N}a_i s_i
  +\sum_{k=1}^{M}
   \log\!\left[
    2\cosh\!\left(b_k+\sum_{i=1}^{N}W_{ik}s_i\right)
   \right].
 \label{eq:rbm-log-amplitude}
\end{equation}
A hidden unit sees all visible spins through the weights $W_{ik}$, so an RBM has a direct way to build nonlocal correlations \cite{CarleoTroyer2017,Lange2024Review}.

The hidden density is
\begin{equation}
 \alpha=\frac{M}{N}.
 \label{eq:hidden-density}
\end{equation}
The model has $N+M+NM$ complex parameters. In the benchmark, $N=8$ and $\alpha=4$, so $M=32$. This gives $296$ complex parameters, or $592$ real trainable numbers.

The RBM amplitude is not normalized automatically. Its Born probability is
\begin{equation}
 p_{\vartheta}(s)
 =\frac{|\widetilde\psi_{\vartheta}(s)|^2}
        {\sum_{s'}|\widetilde\psi_{\vartheta}(s')|^2}.
 \label{eq:rbm-normalization}
\end{equation}
For $N=8$, the denominator contains only $2^8=256$ configurations and can be evaluated exactly. At large $N$, the sum is exponentially large, so an RBM calculation normally uses Monte Carlo sampling. This difference is why the small RBM result in this paper is a clean representation and optimization check, not a large-system scaling result.

\subsection{Autoregressive factorization}

Every joint distribution can be written by the probability chain rule,
\begin{equation}
 q_{\varphi}(x_1,\ldots,x_N)
 =\prod_{j=1}^{N}q_{\varphi}(x_j\mid x_{<j}),
 \qquad
 x_{<j}=(x_1,\ldots,x_{j-1}).
 \label{eq:ar-factorization}
\end{equation}
If each conditional distribution sums to one, then the full distribution is normalized. This can be shown by summing the last variable, then the next-to-last variable, and continuing to the first.

The factorization also gives direct ancestral sampling. First sample $x_1$ from $q(x_1)$. Then sample $x_2$ from $q(x_2\mid x_1)$, and continue until $x_N$. A new ancestral pass gives a new independent sample; no Markov-chain burn-in is required. The log probability is additive,
\begin{equation}
 \log q_{\varphi}(x)
 =\sum_{j=1}^{N}\log q_{\varphi}(x_j\mid x_{<j}).
 \label{eq:ar-log-probability}
\end{equation}
This makes maximum-likelihood training straightforward \cite{HibatAllah2020,Sharir2020}.

A complex autoregressive wavefunction can use the same idea,
\begin{equation}
 \psi_{\vartheta}(x)
 =\prod_{j=1}^{N}
 \left[
 \sqrt{q_{\vartheta}(x_j\mid x_{<j})}
 e^{i\phi_{\vartheta}(x_j,x_{<j})}
 \right].
 \label{eq:complex-ar-wavefunction}
\end{equation}
Local normalization,
\begin{equation}
 \sum_{x_j\in\{0,1\}}q_{\vartheta}(x_j\mid x_{<j})=1,
 \label{eq:local-ar-normalization}
\end{equation}
normalizes the magnitude of the full wavefunction. The phase outputs still have to be learned from the energy or from measurements in enough different bases.

\subsection{GRU hidden-state update}

The positive data-driven model uses a gated recurrent unit (GRU). At site $j$, the previous bit is mapped to an embedding vector $e_j$, and the GRU updates a hidden vector $h_j$. A standard GRU update can be written as
\begin{align}
 r_j&=\sigma(W_r e_j+U_r h_{j-1}+b_r),
 \label{eq:gru-reset}\\
 z_j&=\sigma(W_z e_j+U_z h_{j-1}+b_z),
 \label{eq:gru-update}\\
 n_j&=\tanh\!\left(W_n e_j+b_n
       +r_j\odot(U_n h_{j-1}+c_n)\right),
 \label{eq:gru-candidate}\\
 h_j&=(1-z_j)\odot n_j+z_j\odot h_{j-1}.
 \label{eq:gru-hidden}
\end{align}
Here $\sigma(u)=1/(1+e^{-u})$ is the logistic function and $\odot$ denotes elementwise multiplication. The reset gate $r_j$ controls how much of the old hidden state enters the candidate $n_j$. The update gate $z_j$ mixes the candidate with the previous hidden state. Different software packages group the bias terms differently, but the role of the gates is the same.

The next-bit logits are
\begin{equation}
 \ell_j=W_o h_j+b_o,
 \qquad
 q_{\varphi}(x_{j+1}=a\mid x_{\le j})
 =\frac{e^{\ell_{j,a}}}{\sum_{b=0}^{1}e^{\ell_{j,b}}}.
 \label{eq:gru-output}
\end{equation}
The softmax makes the two probabilities positive and normalized. A beginning-of-sequence token is used before the first physical bit.

During training, the true previous bit is supplied at each step, a procedure known as teacher forcing. A single recurrent pass through a training string therefore provides the conditional probability for every next-bit target. During generation, the true continuation is unavailable, so each sampled bit is fed back as the input for the next step.

The hidden state is not meant to memorize the full prefix. It should keep only the part that matters for future predictions. A running parity is the simplest example: an arbitrarily long bit string can be summarized by one bit. This is closely related to the two-dimensional virtual memory that appears in the cluster-state MPS.

\subsection{Probability-learning loss and evaluation metrics}

For $M$ independent training strings $x^{(\mu)}$, the negative log likelihood per site is
\begin{equation}
 \mathcal L_{\mathrm{NLL}}(\varphi)
 =-\frac{1}{MN}
  \sum_{\mu=1}^{M}\log q_{\varphi}\!\left(x^{(\mu)}\right).
 \label{eq:nll-loss}
\end{equation}
Here $\varphi$ denotes the parameters of the positive probability model. The expected negative log likelihood separates as
\begin{equation}
 \E_{x\sim p}[-\ln q_{\varphi}(x)]
 =H_{\ln}(p)+D_{\mathrm{KL}}(p\Vert q_{\varphi}),
 \label{eq:cross-entropy-decomposition}
\end{equation}
where $H_{\ln}(p)=-\sum_xp(x)\ln p(x)$ is the target entropy in natural-log units. Since this entropy does not depend on the model, minimizing the likelihood loss is the same as minimizing the forward KL divergence. I calculate it in bits per site,
\begin{equation}
 d_{\mathrm{KL}}^{\rightarrow}
 =\frac{1}{N\ln 2}
  \E_{x\sim p}\!\left[\ln\frac{p(x)}{q_{\varphi}(x)}\right].
 \label{eq:forward-kl}
\end{equation}
It is zero only when $q_{\varphi}=p$ on the support of $p$.

The classical fidelity used here is the squared Bhattacharyya coefficient,
\begin{equation}
 F_{\mathrm{cl}}(p,q)
 =\left(\sum_x\sqrt{p(x)q(x)}\right)^2.
 \label{eq:classical-fidelity}
\end{equation}
It lies between zero and one, with one meaning identical distributions. For $N=64$ I use the identity
\begin{equation}
 F_{\mathrm{cl}}(p,q)
 =\left(\E_{x\sim p}\sqrt{\frac{q(x)}{p(x)}}\right)^2
 \label{eq:fidelity-mc-estimator}
\end{equation}
and estimate the expectation from exact-MPS Monte Carlo samples. This avoids summing all $2^{64}$ strings. The estimator is reliable only when the sampled target support also covers the regions where $q$ places substantial weight, which is why I consider KL and conditional errors as well.

Because the exact MPS sampler also gives the target conditional probability $p(x_j=1\mid x_{<j})$, I calculate
\begin{equation}
 \mathrm{MAE}_{\mathrm{cond}}
 =\frac{1}{N}\sum_{j=1}^{N}
 \E_{x\sim p}
 \left|q_{\varphi}(x_j=1\mid x_{<j})
       -p(x_j=1\mid x_{<j})\right|.
 \label{eq:conditional-mae}
\end{equation}
This metric asks whether the hidden state predicts the correct next-site rule, not only whether the total log likelihood is good.

For samples generated from the learned model, I also estimate the reverse KL divergence,
\begin{equation}
 d_{\mathrm{KL}}^{\leftarrow}
 =\frac{1}{N\ln2}
  \E_{x\sim q_{\varphi}}
  \left[\ln\frac{q_{\varphi}(x)}{p(x)}\right].
 \label{eq:reverse-kl}
\end{equation}
The forward and reverse directions weight errors differently, so reporting both gives a better check for missed or overproduced regions of the distribution.

Finally, I compare simple observables. For bit $x_j$, the corresponding Pauli-$Z$ value is $z_j=1-2x_j\in\{+1,-1\}$. The site magnetization and connected correlation are
\begin{equation}
 m_j=\E[z_j],
 \qquad
 C_{jk}=\E[z_jz_k]-\E[z_j]\E[z_k].
 \label{eq:classical-observables}
\end{equation}
Root-mean-square errors over sites or site pairs measure how well samples from the learned model reproduce these quantities.

\subsection{Autoregressive models compared with an RBM}

The autoregressive form has three practical advantages over an RBM. Its normalization is local, its likelihood is explicit, and it can sample directly in one left-to-right pass. These features remove the partition-function estimate and can remove MCMC autocorrelation.

The RBM has a different advantage: a hidden unit connects to the whole visible layer, so the model does not have to transport information through the chain step by step. It can build global correlations directly. The cost is a global normalization and, at large size, usually a Monte Carlo sampler.

The main weakness of a generic RNN is not necessarily hidden-state size. It is that the network must learn what the hidden coordinates mean and how phase information should update. For the rotated cluster state, the exact answer is already known to use a two-dimensional MPS bond. The MPS-induced recurrent model in \cref{sec:mps-induced-model} builds that update into the architecture. It is still trained from random parameters, but it does not have to invent the form of the memory.

The generic variational RNN used here is not this autoregressive model. It is a tanh RNN that returns a complex log-amplitude directly and uses local Monte Carlo sampling. Its samples can be correlated during these VMC runs, so its updates are not directly comparable with the deterministic MPS calculation.

%% file: sections/03_mps.tex
\section{Matrix product states and the MPS-induced recurrent model}
\label{sec:mps}

MPS are used in two ways in this paper. First, they provide exact reference states for the cluster and AKLT calculations. Second, the same left-to-right contraction is turned into a trainable recurrent ansatz. This section introduces the required MPS formulas before using them in the experiments.

\subsection{Schmidt decomposition, bond dimension, and MPS form}

For any cut that divides a chain into a left part $L$ and a right part $R$, a pure state has a Schmidt decomposition
\begin{equation}
 \ket{\psi}
 =\sum_{\alpha=1}^{D_{L|R}}
 \lambda_{\alpha}
 \ket{\alpha}_{L}\ket{\alpha}_{R}.
 \label{eq:schmidt-decomposition}
\end{equation}
The Schmidt coefficients $\lambda_{\alpha}$ are non-negative, and $D_{L|R}$ is the number of nonzero coefficients. This number is the Schmidt rank across the cut. It measures how many independent left states must be paired with right states.

Applying the Schmidt decomposition successively along an open chain gives an MPS \cite{PerezGarcia2007,Schollwock2011,Orus2014}. Its amplitudes are
\begin{equation}
 \psi(x_1,\ldots,x_N)
 =l^{\dagger}A_1^{x_1}A_2^{x_2}\cdots A_N^{x_N}r.
 \label{eq:mps-amplitude}
\end{equation}
At site $j$, the physical outcome $x_j$ selects one matrix $A_j^{x_j}$. The boundary vectors $l$ and $r$ close the open chain, and $l^{\dagger}$ is the conjugate-transpose row vector. The largest internal matrix size is the bond dimension $D$. An exact MPS can be chosen with $D$ equal to the largest Schmidt rank over all cuts.

The contraction can be read as a recurrence. After the first $j$ sites,
\begin{equation}
 v_j^{\dagger}=l^{\dagger}A_1^{x_1}\cdots A_j^{x_j},
 \qquad
 v_{j+1}^{\dagger}=v_j^{\dagger}A_{j+1}^{x_{j+1}}.
 \label{eq:mps-recurrence}
\end{equation}
The vector $v_j$ is the virtual information passed to the next site. It has only $D$ complex components even though the prefix $x_{1:j}$ can be arbitrarily long. This is why an MPS can be interpreted as a linear recurrent model: the hidden state is the bond vector and the observed bit chooses the update matrix \cite{Wu2023}.

A local unitary on one physical site changes the physical basis of its tensor but not the Schmidt coefficients across any cut. Therefore the rotation in \cref{eq:rotated-H-state} cannot increase or decrease the exact Schmidt rank. The cluster state remains a $D=2$ MPS at every angle.

\subsection{Transfer channel and ordinary correlations}
\label{sec:transfer-channel}

Probabilities and expectation values contain a tensor and its complex conjugate. It is convenient to combine these two layers into a transfer channel. For a site tensor $A^x$, define
\begin{equation}
 \mathcal E(X)=\sum_x A^x X A^{x\dagger}.
 \label{eq:transfer-channel}
\end{equation}
The matrix $X$ represents the uncontracted virtual part to the right. Applying $\mathcal E$ adds one physical site and sums over its outcome. In a translation-invariant infinite MPS, the dominant left and right fixed points of this channel provide the stationary boundaries.

I use $\spec(\mathcal E)$ for the set of eigenvalues of a transfer channel and $\rank(\mathcal E)$ for its matrix rank. If the leading transfer eigenvalue is normalized to $\lambda_1=1$ and the next relevant eigenvalue is $\lambda_2$, a usual exponential correlation length is
\begin{equation}
 \xitwo=-\frac{1}{\ln|\lambda_2|}.
 \label{eq:mps-correlation-length}
\end{equation}
This formula follows because repeated transfer steps produce factors $\lambda_2^r=e^{-r/\xitwo}$. If $\lambda_2=0$, the corresponding mode disappears after a finite number of steps rather than having a long exponential tail.

A physical basis change modifies each outcome tensor coherently. For a one-site unitary $U$,
\begin{equation}
 \widetilde A^{x}=\sum_y\bra{x}U\ket{y}A^y.
 \label{eq:rotated-tensor}
\end{equation}
Unitarity gives
\begin{equation}
 \sum_x\widetilde A^x X\widetilde A^{x\dagger}
 =\sum_y A^y X A^{y\dagger}.
 \label{eq:channel-invariance}
\end{equation}
Thus the outcome-averaged transfer channel is unchanged by a local basis rotation, even though the individual conditioned maps $\widetilde A^x$ change. This is central to the CMI calculation: ordinary averaged correlations can stay short while a particular measurement record preserves virtual information for a long time.

\subsection{Exact MPS probabilities and sequential sampling}
\label{sec:mps-sampling}

For a finite chain, I precompute right environments. Let
\begin{equation}
 R_N=rr^{\dagger},
 \qquad
 R_j=\sum_{x_{j+1}}
 A_{j+1}^{x_{j+1}}R_{j+1}A_{j+1}^{x_{j+1}\dagger}
 \quad (j=N-1,\ldots,0).
 \label{eq:right-environment}
\end{equation}
The matrix $R_j$ is the contraction of sites $j+1$ through $N$, with their physical outcomes summed out. After a prefix $x_{1:j}$ has produced the left vector $v_j$ in \cref{eq:mps-recurrence}, the exact next-site probability is
\begin{equation}
 p(x_{j+1}=a\mid x_{1:j})
 =\frac{
 v_j^{\dagger}A_{j+1}^{a}R_{j+1}A_{j+1}^{a\dagger}v_j
 }{
 v_j^{\dagger}R_jv_j
 }.
 \label{eq:mps-conditional-sampling}
\end{equation}
The numerator fixes the next outcome to $a$ and sums the remaining sites. The denominator sums both possible next outcomes, so the two conditional probabilities add to one.

Repeating \cref{eq:mps-conditional-sampling} from left to right gives an independent exact sample. The product of the selected conditionals is
\begin{equation}
 \prod_{j=1}^{N}p(x_j\mid x_{<j})=|\psi(x)|^2.
 \label{eq:mps-probability-product}
\end{equation}
I use this equality as an internal sampler check by comparing the sequential product with a direct contraction of the completed string.

The same contractions give the joint distributions needed for CMI. For a finite endpoint protocol, I contract from both physical boundaries. For a stationary bulk protocol, I use the fixed points of the infinite-chain transfer channel as effective boundaries. In both cases, the resulting probabilities are checked for normalization and non-negativity before \cref{eq:cmi-probability} is evaluated.

\subsection{Cluster MPS and the two readout protocols}
\label{sec:cluster-readouts}

Let
\begin{equation}
 U_H=H_1H_2\cdots H_{N-1},
 \qquad
 H=\frac{1}{\sqrt2}\begin{pmatrix}1&1\\1&-1\end{pmatrix}
 \label{eq:hadamard-product}
\end{equation}
be a product of Hadamard gates. The Hamiltonians in \cref{eq:HC-intro,eq:HES-intro} satisfy
\begin{equation}
 H_{\mathrm{ES}}=U_H H_{\mathrm C}U_H^{\dagger}.
 \label{eq:cluster-unitary-relation}
\end{equation}
They therefore have the same spectrum and entanglement, but their computational-basis data correspond to different physical measurements.

A convenient bulk tensor for the conventional graph-cluster state is
\begin{equation}
 A^0=\frac{1}{\sqrt2}
 \begin{pmatrix}1&0\\1&0\end{pmatrix},
 \qquad
 A^1=\frac{1}{\sqrt2}
 \begin{pmatrix}0&1\\0&-1\end{pmatrix}.
 \label{eq:cluster-mps-tensor}
\end{equation}
Its double-layer transfer spectrum is
\begin{equation}
 \spec(\mathcal E_{\mathrm C})=\{1,0,0,0\},
 \qquad
 \rank\mathcal E_{\mathrm C}=2,
 \qquad
 \rank\mathcal E_{\mathrm C}^{2}=1.
 \label{eq:cluster-transfer-spectrum}
\end{equation}
All traceless averaged components disappear after at most two bulk transfer steps. The conventional two-point process therefore has only a short finite transient. By \cref{eq:channel-invariance}, rotating the physical basis changes conditioned records but not this averaged channel.

At $\theta=0$, the entanglement-swapping readout has a deterministic endpoint parity rule,
\begin{equation}
 x_1\oplus x_N=x_3\oplus x_5\oplus\cdots\oplus x_{N-1},
 \label{eq:hes-parity}
\end{equation}
where $\oplus$ is addition modulo two. Once the middle record is known, one unbiased endpoint determines the other. Hence
\begin{equation}
 I(X_1:X_N\mid X_2,\ldots,X_{N-1})=1\ \text{bit}.
 \label{eq:hes-one-bit}
\end{equation}
At $\theta=\pi/2$, the endpoint CMI is zero. Both limits are obtained directly from exact MPS probabilities.

The conventional graph-cluster state gives a useful contrast. It can be written as
\begin{equation}
 \ket{C_N}=\prod_{j=1}^{N-1}CZ_{j,j+1}\ket{+}^{\otimes N},
 \qquad
 \ket{+}=\frac{\ket0+\ket1}{\sqrt2}.
 \label{eq:graph-cluster-state}
\end{equation}
Here $CZ=\operatorname{diag}(1,1,1,-1)$ is the controlled-$Z$ gate. Every computational-basis amplitude has magnitude $2^{-N/2}$, because controlled-$Z$ gates only add signs. The $Z$-basis distribution is therefore uniform at $\theta=0$, and its $Z$-basis CMI is zero. The one-bit relation in \cref{eq:hes-one-bit} comes from the readout protocol and boundary choice, not from a larger bond dimension.

\subsection{CMI scan: design, fitting, and result}
\label{sec:cmi-scan}

The two cluster scans use the same rotation grid but different operational partitions.

For the conventional cluster state, I use a stationary bulk block and define
\begin{equation}
 I_m^{\mathrm C}(\theta)
 =I(X_0:X_{m+1}\mid X_1,\ldots,X_m).
 \label{eq:cluster-stationary-cmi}
\end{equation}
The probability block is generated from infinite-MPS fixed points. I evaluate $m=0,\ldots,12$ on $31$ equally spaced angles between $0$ and $\pi/2$. The MPS contraction for one fixed string is cheap, but the full joint table contains $2^{m+2}$ strings. This exponential table size, rather than the bond contraction itself, sets the practical range of $m$.

For the entanglement-swapping protocol, I use the physical endpoints of a finite chain,
\begin{equation}
 I_m^{\mathrm{ES}}(\theta)
 =I(X_1:X_{m+2}\mid X_2,\ldots,X_{m+1}),
 \label{eq:hes-finite-cmi}
\end{equation}
where the buffer length $m=N-2$ is even. I evaluate $m=2,4,\ldots,14$ on the same angle grid. Increasing $m$ also increases the chain length, so this is a finite endpoint sequence rather than a stationary bulk quantity.

For the HES protocol, the known upper bound \cite{Yang2024}
\begin{equation}
 I_m^{\mathrm{ES}}(\theta)\le |\cos\theta|^m
 \label{eq:yang-bound}
\end{equation}
has the exponential envelope length
\begin{equation}
 \xi_{\mathrm{bound}}(\theta)
 =-\frac{1}{\ln|\cos\theta|}.
 \label{eq:yang-bound-length}
\end{equation}
This is an upper-bound scale, not an exact CMI length.

The first analysis used one log-linear fit at each angle. This gave very large length estimates near angles where the CMI was already very small. I therefore changed the analysis in three ways. First, for the stationary protocol, even and odd buffer lengths are considered separately. Second, I check the effective length in \cref{eq:cmi-effective-length}. An exponential decay should give an approximately constant value over the fitting range. Third, I do not use a fitted length when the CMI is too small or when the even- and odd-length results differ strongly. The CMI amplitude is shown together with the fitted length to make this comparison clear.

For the stationary cluster calculation, I use a finite-window fit only when these checks give a reasonable result. For HES, I fit $m=4,6,\ldots,14$ and use the result only as a summary of this finite-size data. These fitted lengths describe the data in the calculated range and do not determine the asymptotic behavior.

\begin{figure}[!htbp]
\centering
\includegraphics[width=0.96\textwidth]{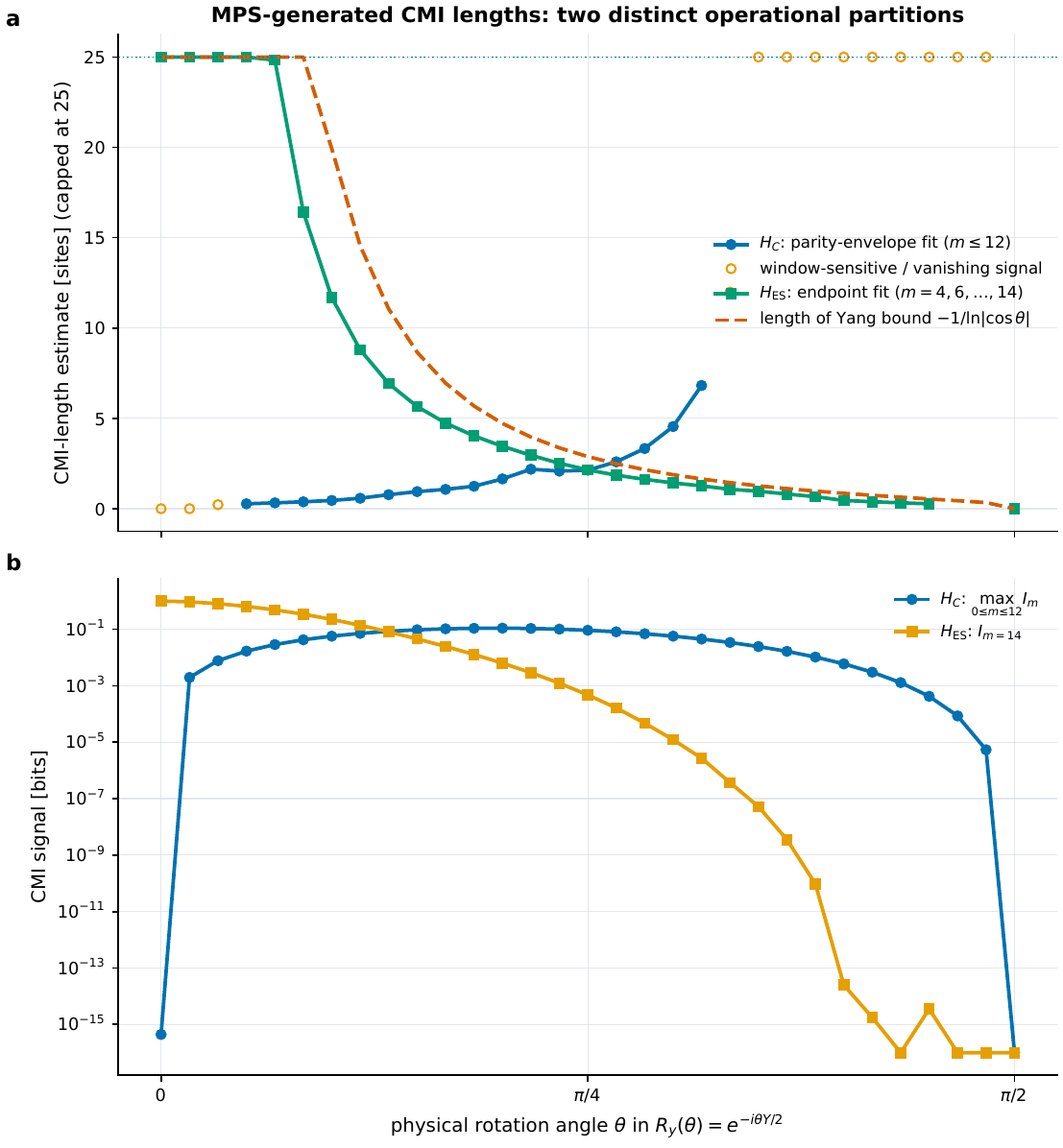}
\caption{CMI from exact MPS probabilities. The upper panel shows finite-window length estimates for the stationary $H_{\mathrm C}$ bulk protocol and the finite-endpoint $H_{\mathrm{ES}}$ protocol. Open conventional-cluster markers indicate fits that are sensitive to the fitting window or have very small CMI. The dashed curve shows the length scale obtained from the upper bound in \cref{eq:yang-bound}. The lower panel shows a representative CMI amplitude, which helps indicate where the fitted length is reliable.}
\label{fig:cluster-cmi-length}
\end{figure}

The HES endpoint sequence changes from one bit of conditional information at $\theta=0$ to zero at $\theta=\pi/2$. At $\theta=0.35$, the direct finite-window fit gives
\begin{equation}
 \xi_{\mathrm{fit}}\simeq9.57,
 \qquad
 \xi_{\mathrm{bound}}\simeq15.99.
 \label{eq:theta035-lengths}
\end{equation}
These values have different meanings: the first is obtained by fitting the exact data over the calculated window, while the second comes from the upper bound.

The stationary conventional-cluster sequence is zero at both endpoint angles and nonzero between them. The two curves in \cref{fig:cluster-cmi-length} are not two estimates of the same quantity. The first comes from stationary bulk variables, while the second uses the physical endpoints of a finite chain whose length changes with $m$. The boundary conditions, measurement basis, and choice of endpoints therefore affect the two quantities.

\subsection{AKLT chain as an exact control}
\label{sec:aklt-control}

The spin-1 Affleck--Kennedy--Lieb--Tasaki (AKLT) chain is a second bond-dimension-two state \cite{AKLT1987,AKLT1988}. Its Hamiltonian is
\begin{equation}
 H_{\mathrm{AKLT}}=\sum_j P_{j,j+1}^{(2)},
 \label{eq:aklt-hamiltonian}
\end{equation}
where $P_{j,j+1}^{(2)}$ projects two neighboring spin-1 particles onto total spin two. Writing $\mathbf S_j=(S_j^x,S_j^y,S_j^z)$ for the three spin-1 operators at site $j$, the projector is
\begin{equation}
 P_{j,j+1}^{(2)}
 =\frac16(\mathbf S_j\!\cdot\!\mathbf S_{j+1})^2
  +\frac12\mathbf S_j\!\cdot\!\mathbf S_{j+1}
  +\frac13\id.
 \label{eq:aklt-projector}
\end{equation}
This polynomial equals one in the total-spin-two sector and zero in the total-spin-zero and total-spin-one sectors.

In the valence-bond picture, each physical spin one is made from two virtual spin-$1/2$ degrees of freedom. Neighboring virtual spins form singlets, and the two virtual spins on each site are projected onto the symmetric triplet. One virtual spin-$1/2$ crosses each bulk cut, so the exact bulk bond dimension is two. The unpaired virtual spins at the ends of an open chain are the Haldane-phase edge modes \cite{Pollmann2012}.

For the translation-invariant infinite state, I use
\begin{equation}
 A^{+}=\sqrt{\frac23}\,\sigma^{+},
 \qquad
 A^{0}=-\frac{1}{\sqrt3}\sigma^z,
 \qquad
 A^{-}=-\sqrt{\frac23}\,\sigma^{-},
 \label{eq:aklt-tensors}
\end{equation}
where $\sigma^{\pm}=(X\pm iY)/2$. The transfer spectrum is
\begin{equation}
 \spec(\mathcal E_{\mathrm{AKLT}})
 =\left\{1,-\frac13,-\frac13,-\frac13\right\}.
 \label{eq:aklt-transfer-spectrum}
\end{equation}
Therefore ordinary spin correlations contain $(-1/3)^r$ and have length
\begin{equation}
 \xitwo=\frac{1}{\ln3}.
 \label{eq:aklt-two-point-length}
\end{equation}

The measured process has an especially simple memory rule. Outcomes $+$ and $-$ select rank-one matrices, so they reset the incoming virtual direction. Outcome $0$ is proportional to the unitary $\sigma^z$ and occurs with probability $1/3$. Endpoint information survives a buffer of length $m$ only if all $m$ outcomes are zero. That record has probability $3^{-m}$.

The remaining factor $4/9$ can be seen from the two adjacent endpoint outcomes. The stationary incoming virtual state is the maximally mixed bond state $\rho=\id/2$. Their joint probability is
\begin{equation}
 P(a,c)=\Tr\!\left[A^cA^a\rho A^{a\dagger}A^{c\dagger}\right].
 \label{eq:aklt-two-site-probability}
\end{equation}
Substituting the tensors in \cref{eq:aklt-tensors} gives the following matrix in the outcome order $(+,0,-)$:
\begin{equation}
 P_{ac}=\frac{1}{9}
 \begin{pmatrix}
 0&1&2\\
 1&1&1\\
 2&1&0
 \end{pmatrix}_{ac}.
 \label{eq:aklt-adjacent-probability}
\end{equation}
Every row and column sums to $1/3$. Five nonzero entries equal $1/9$ and give no logarithmic contribution to mutual information, while two entries equal $2/9$ and contribute $\log_2 2=1$. Therefore
\begin{equation}
 I(S_0:S_1)=2\left(\frac{2}{9}\right)=\frac{4}{9}\ \text{bits}.
 \label{eq:aklt-adjacent-mi}
\end{equation}
An all-zero buffer changes the virtual state only by a known Pauli-$Z$ byproduct, so this adjacent mutual information is unchanged. Every other buffer record contains a rank-one reset and contributes zero. Averaging over records therefore gives
\begin{equation}
 I(S_0:S_{m+1}\mid S_1,\ldots,S_m)
 =\frac49\,3^{-m}\ \text{bits}.
 \label{eq:aklt-cmi-law}
\end{equation}
Comparing \cref{eq:aklt-cmi-law} with \cref{eq:cmi-exponential} gives
\begin{equation}
 \xicmi=\frac{1}{\ln3}=\xitwo\simeq0.910.
 \label{eq:aklt-equal-lengths}
\end{equation}
The two lengths agree because the same factor $1/3$ is both the magnitude of the subleading transfer eigenvalue and the probability of the only outcome that preserves the virtual memory. There is no general identity of this kind.

The infinite AKLT state is spin-rotation invariant. Rotating every measurement axis changes the virtual basis and outcome labels but not the distribution. Exact MPS contractions reproduce \cref{eq:aklt-cmi-law} to numerical precision for the tested angles and buffer lengths.

\subsection{MPS-induced recurrent model at 64 sites}
\label{sec:mps-induced-model}

\subsubsection{Architecture and capacity control}

The trainable model uses separate boundary tensors and a two-site bulk cell,
\begin{equation}
 \psi_{\Theta}(x_1,\ldots,x_N)
 =L^{x_1}A_{\mathrm{even}}^{x_2}A_{\mathrm{odd}}^{x_3}
 \cdots R^{x_N}.
 \label{eq:structured-mps-ansatz}
\end{equation}
The two bulk tensors allow the recurrent rule to alternate between the two sublattices, matching the alternating $X/Z$ structure of the HES representation without fixing the tensors to their exact values. For an internal bond dimension $\chi$, $L^x$ is a $1\times\chi$ row, $R^x$ is a $\chi\times1$ column, and the two bulk tensors are $\chi\times\chi$ matrices. The recurrence is
\begin{equation}
 h_1=L^{x_1},
 \qquad
 h_j=h_{j-1}A_{\pi(j)}^{x_j},
 \qquad
 \psi_{\Theta}(x)=h_{N-1}R^{x_N},
 \label{eq:structured-recurrence}
\end{equation}
where $\pi(j)$ selects the even or odd tensor. Mathematically this is an MPS, closely related to the tensorial recurrent networks of Ref.~\cite{Wu2023}. I call it MPS-induced recurrent because its forward pass is a recurrent hidden-state update. Its role in this paper is a structure-matched positive control rather than a new architecture.

The number of complex parameters is $4\chi+4\chi^2$, or $8\chi+8\chi^2$ real parameters. Thus $\chi=2$ has only $48$ real raw parameters. The exact rotated cluster state belongs to this family. The $\chi=1$ version has $16$ real parameters and is a product state. It cannot carry a nontrivial virtual parity, so it is a direct negative control for representational capacity.

\subsubsection{Optimization design and improvements}

The target is $N=64$ and $\theta=0.35$. All four tensor groups start from independent complex Gaussian entries. After each Adam step I rescale the Frobenius norm of each tensor group. This does not insert the exact solution. It only removes a simple numerical freedom: one tensor group can grow while another shrinks, leaving the normalized wavefunction almost unchanged but making the raw parameters badly scaled. The rescaling made the optimization much more stable.
For example, two neighboring tensor groups can be rescaled as
\begin{equation}
 A^{x}_{\mathrm{even}} \rightarrow c A^{x}_{\mathrm{even}},
 \qquad
 A^{x}_{\mathrm{odd}} \rightarrow c^{-1} A^{x}_{\mathrm{odd}},
\end{equation}
so that their product is unchanged,
\begin{equation}
 (c A^{x}_{\mathrm{even}})
 (c^{-1} A^{x}_{\mathrm{odd}})
 =
 A^{x}_{\mathrm{even}} A^{x}_{\mathrm{odd}}.
\end{equation}
Thus the same normalized wavefunction can correspond to tensor parameters with very different numerical scales.

The norm, total energy, and each individual stabilizer expectation are evaluated by exact double-layer contraction. Prefix and suffix environments are reused, so the cost of one energy evaluation grows linearly with $N$ at fixed $\chi$. There is no MCMC sampler and no stochastic gradient noise.

I first run $150$ Adam steps with learning rate $10^{-2}$ and gradient clipping at $10$. Adam reliably finds the correct basin, but it typically stops with a total error of order $10^{-5}$ to $10^{-4}$. I then run at most $50$ L-BFGS iterations with a strong-Wolfe line search. This two-stage schedule was kept because it was both seed-robust and easy to diagnose.

Five independent $\chi=2$ initializations are trained. The $\chi=1$ control uses the same Hamiltonian and optimizer schedule. This comparison changes only the virtual capacity.

\subsubsection{Result}

After Adam, the five $\chi=2$ total errors lie between $1.23\times10^{-5}$ and $2.82\times10^{-4}$. After L-BFGS,
\begin{equation}
 \operatorname{median}_{\mathrm{seeds}}(E-E_0)=9.98\times10^{-12},
 \label{eq:structured-total-median}
\end{equation}
with a full range from $3.98\times10^{-13}$ to $1.07\times10^{-10}$. The median energy-density error is
\begin{equation}
 \operatorname{median}_{\mathrm{seeds}}(\epsilon_E)=1.56\times10^{-13}.
 \label{eq:structured-density-result}
\end{equation}
Every $\chi=2$ initialization reaches the ground-state energy to high precision.

Because the Hamiltonian is frustration free,
\begin{equation}
 E-E_0=\sum_{j=1}^{N}\left(1-\langle K_j\rangle\right).
 \label{eq:stabilizer-error-sum}
\end{equation}
A small total error could in principle hide one bad boundary term among many good bulk terms, so I also inspect every $\langle K_j\rangle$. The smallest value over all sites and all seeds is
\begin{equation}
 \min_{j,\mathrm{seeds}}\langle K_j\rangle
 =0.999999999956.
 \label{eq:min-stabilizer-result}
\end{equation}
This confirms that the final energy is not produced by cancellation between local errors.

The $\chi=1$ control remains at
\begin{equation}
 E-E_0=32,
 \qquad
 \frac1N\sum_j\langle K_j\rangle=\frac12.
 \label{eq:chi1-result}
\end{equation}
The optimizer is working, but the product-state family lacks the required bond memory.

\begin{figure}[!htbp]
\centering
\includegraphics[width=0.90\textwidth]{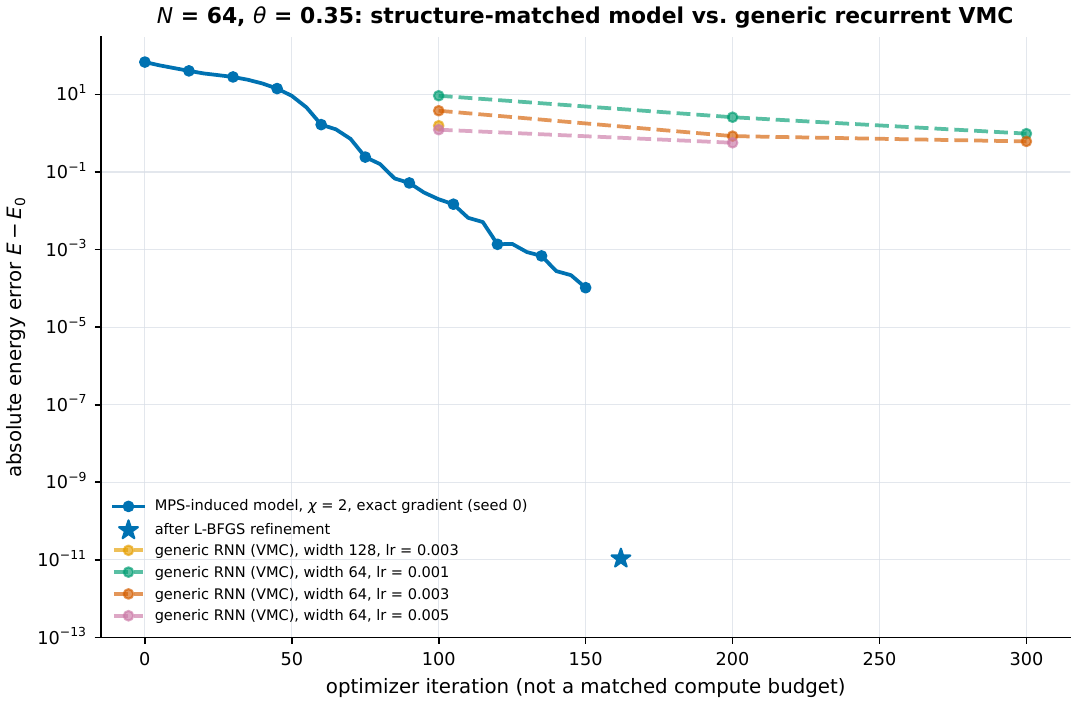}
\caption{Optimization at $N=64$ and $\theta=0.35$. The solid curve is one $\chi=2$ MPS-induced recurrent run, and the star is the final L-BFGS value. Dashed curves are the incomplete generic recurrent VMC runs from \cref{sec:generic-rnn-results}. The update counts do not represent matched computational cost.}
\label{fig:structured-convergence}
\end{figure}

 The MPS-induced model begins from random parameters, yet it already knows the form of the memory: a two-component complex bond vector updated by a physical-outcome-dependent matrix. The generic recurrent ansatz must learn an equivalent coordinate and phase rule from noisy energy estimates. The RBM has global hidden units, but it also lacks the explicit two-dimensional bond update. For this cluster-state problem, the correct inductive bias appears to be a much larger advantage than adding generic hidden units. This comparison, however, changes two things at once: the architecture and the gradient source, which is exact here and sampled in the generic VMC runs. Training the same $\chi=2$ model with the VMC sampler of \cref{sec:generic-rnn-results} would separate the two effects and is left for future work.

\subsection{Independent zero-angle and exact-state checks}
\label{sec:mps-checks}

At $\theta=0$, I compare the analytic stabilizer energy $E_0=-N$ with four numerical routes: exact $D=2$ MPS contraction at $N=64$, sparse exact diagonalization at $N=12$, the full-basis RBM at $N=8$, and the trainable $\chi=2$ recurrent MPS at $N=64$. The system sizes and algorithms differ, so this is a consistency check rather than a timing comparison.

For the trainable $\chi=2$ model, the median zero-angle energy-density error over five seeds is
\begin{equation}
 6.57\times10^{-14},
 \label{eq:structured-zero-angle}
\end{equation}
with full range $[2.44\times10^{-15},5.77\times10^{-13}]$. The exact $D=2$ contraction reproduces the analytic stabilizer energy $E_0=-N$ to machine precision.

\begin{figure}[!htbp]
\centering
\includegraphics[width=0.96\textwidth]{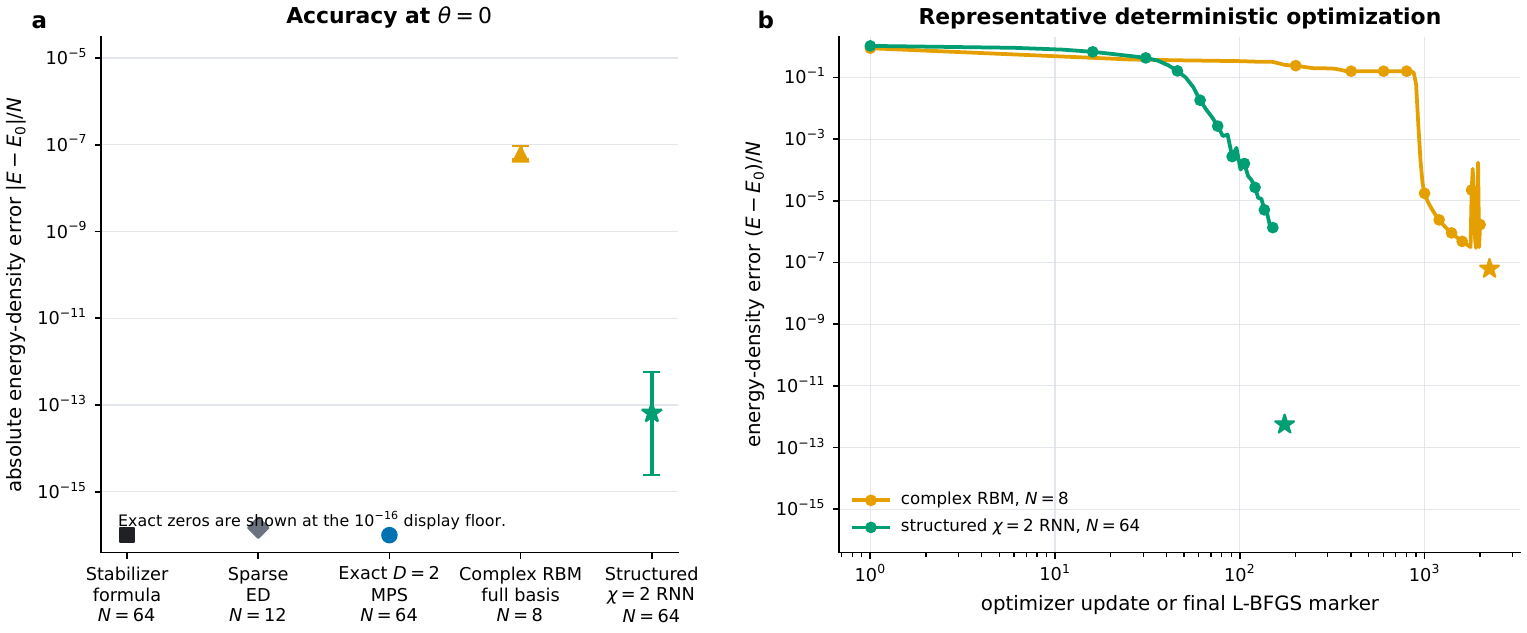}
\caption{Independent checks at $\theta=0$. Panel (a) compares final energy-density errors. Exact zeros are shown at a $10^{-16}$ display floor. Panel (b) shows one deterministic optimization trajectory for the full-basis RBM and one for the MPS-induced recurrent model; the star marks the L-BFGS-refined result.}
\label{fig:theta0-benchmark}
\end{figure}

I also compare the explicit HES MPS with sparse exact ground states for $N=4,6,8$. The absolute overlap is one to numerical precision, the energy is $-N$, and the largest stabilizer residual is $1.11\times10^{-16}$. These checks, together with the conditional-probability check in \cref{eq:mps-probability-product}, the endpoint results in \cref{eq:hes-one-bit}, and the AKLT result in \cref{eq:aklt-cmi-law}, verify the MPS construction, normalization, sampling, and CMI calculation.

%% file: sections/04_rbm_rnn_results.tex
\section{RBM and recurrent neural network results}
\label{sec:neural-results}

In this section, the RBM and variational RNN try to learn a complex ground-state wavefunction. The data-driven GRU learns only the computational-basis probability distribution.

The large-system experiments use the same target, $N=64$ and $\theta=0.35$. The size $N=64$ is far beyond full-basis enumeration but still cheap for an exact bond-dimension-two MPS contraction. The angle is deliberately away from both simple limits: at $\theta=0$ the HES endpoint parity is exact, while at $\theta=\pi/2$ the endpoint memory is erased. At $\theta=0.35$ the measured process has a long but finite conditional memory, so the learning task is nontrivial without changing the exact bond dimension.

\subsection{Complex RBM: exact small-system benchmark}
\label{sec:rbm-results}

\subsubsection{RBM architecture}

The RBM experiment asks whether a generic complex neural ansatz can represent the cluster ground state when sampling noise is removed. I choose $H_{\mathrm{ES}}(0)$ at $N=8$. The exact energy is $E_0=-8$, and the Hilbert space contains only $256$ basis states. Every amplitude, normalization factor, energy, and gradient can therefore be calculated by full-basis summation.

The RBM uses hidden density $\alpha=4$, giving $M=32$ hidden units and $592$ real parameters. Each complex parameter is initialized with independent real and imaginary Gaussian parts of scale $10^{-2}$. I repeat the optimization for five seeds. Since there is no Monte Carlo sampler, differences between seeds come from the non-convex optimization landscape rather than statistical sampling.

I use two optimization stages. The first stage is $2000$ Adam updates with learning rate $2\times10^{-2}$ and gradient clipping. Adam is useful while the parameters are still far from a minimum. The second stage is L-BFGS with a strong-Wolfe line search. L-BFGS uses curvature information from the deterministic objective and gives a cleaner final refinement. This two-stage schedule was kept because Adam alone approached the ground state quickly but left a visible small residual.

\subsubsection{Result}

The median final total energy error over five seeds is
\begin{equation}
 \operatorname{median}(E-E_0)=4.83\times10^{-7},
 \label{eq:rbm-total-result}
\end{equation}
and the median energy-density error is
\begin{equation}
 \operatorname{median}(\epsilon_E)=6.03\times10^{-8}.
 \label{eq:rbm-density-result}
\end{equation}
The full energy-density range is $[4.56,9.28]\times10^{-8}$. These values show that the complex RBM can represent and optimize this small cluster state accurately.

The design also has a limitation. Full-basis summation grows as $2^N$, so the same experiment is not a practical route to $N=64$. A sampled large-system RBM would introduce MCMC noise, autocorrelation, and a different optimization problem. I therefore use the RBM as an exact small-system benchmark and do not extrapolate its error to large $N$.

\subsection{Data-driven autoregressive GRU}
\label{sec:data-driven-results}

\subsubsection{Data generation}

The second experiment asks a different question: can a small recurrent model learn the measured distribution from a finite number of strings, without seeing the Hamiltonian or the exact wavefunction?

The target is the $N=64$ rotated $H_{\mathrm{ES}}$ ground state at
\begin{equation}
 \theta=0.35.
 \label{eq:data-target}
\end{equation}
Independent computational-basis strings are drawn from the exact bond-dimension-two MPS sampler described in \cref{sec:mps-sampling}. The MPS is used to generate data and to evaluate target probabilities, but it is not queried during a gradient update. The GRU therefore has to infer the conditional rule from examples.

I use three training-set sizes,
\begin{equation}
 M\in\{512,2048,8192\},
 \label{eq:data-budgets}
\end{equation}
with three random initializations for each size. Each data budget is four times the previous one, which makes the data-scaling trend easy to read without requiring many separate runs. The validation and test sets contain $4096$ and $8192$ fresh target strings. Keeping them independent prevents training examples from being reused as evaluation data.

The model has an embedding dimension of $8$, a hidden dimension of $16$, and a two-logit output layer. It contains $1306$ real parameters. I keep this network intentionally small; the point is to test whether a generic recurrent state can compress the useful history, not to win by making the network very wide. Training minimizes \cref{eq:nll-loss} with AdamW, learning rate $3\times10^{-3}$, batch size $512$, and gradient clipping at one. A run lasts at most $35$ epochs. The parameters from the best validation epoch are retained, and training stops early after seven epochs without improvement. Early stopping was added because the smallest data set began to overfit before the maximum epoch count.

\subsubsection{Results and effect of more data}

\begin{figure}[!htbp]
\centering
\includegraphics[width=0.96\textwidth]{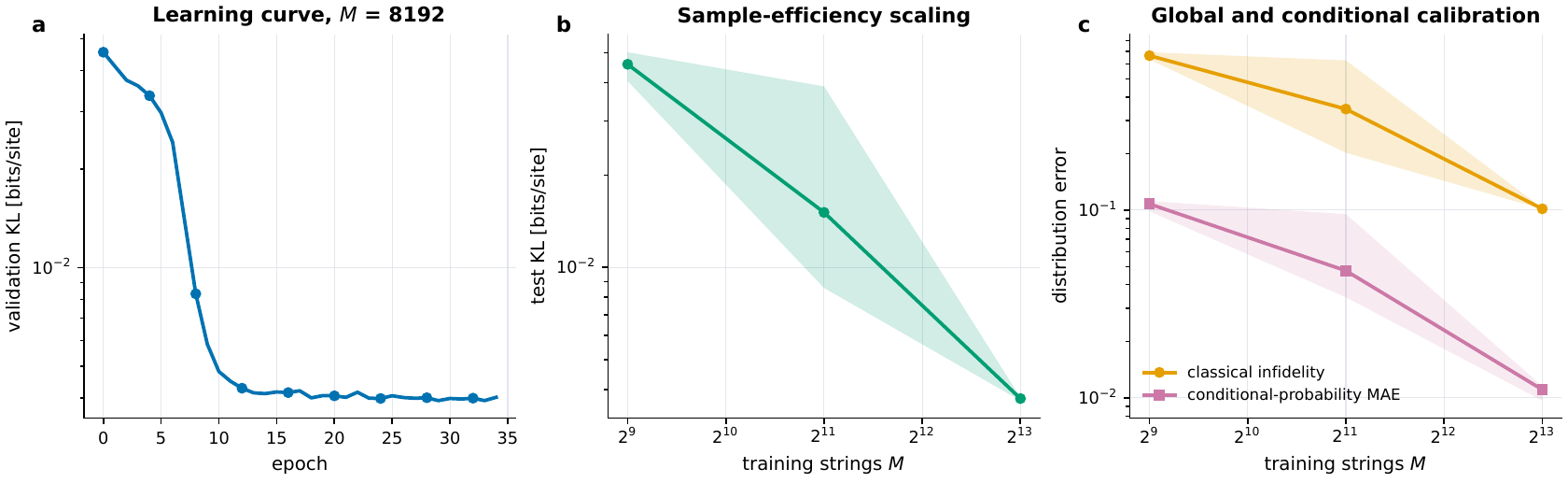}
\caption{Learning the $N=64$, $\theta=0.35$ computational-basis distribution. Panel (a) shows one validation-KL curve for $M=8192$. Panel (b) shows the test forward KL as the training set grows. Panel (c) shows classical infidelity $1-F_{\mathrm{cl}}$ and conditional-probability error. Lines are medians over three seeds and shaded regions show the full seed range.}
\label{fig:data-driven-scaling}
\end{figure}

\begin{table}[!htbp]
\centering
\small
\caption{Held-out test results for the data-driven GRU. Values are medians with the full $[\min,\max]$ range over three seeds.}
\label{tab:data-driven-results}
\begin{tabular}{@{}cccc@{}}
\toprule
Training strings $M$ & $d_{\mathrm{KL}}^{\rightarrow}$ [bits/site] & $F_{\mathrm{cl}}$ & conditional MAE \\
\midrule
$512$
& $4.59\times10^{-2}$ $[4.05,5.03]\times10^{-2}$
& $0.335$ $[0.307,0.365]$
& $0.108$ $[0.0985,0.1117]$ \\
$2048$
& $1.51\times10^{-2}$ $[8.58\times10^{-3},3.89\times10^{-2}]$
& $0.655$ $[0.373,0.798]$
& $0.0475$ $[0.0343,0.0951]$ \\
$8192$
& $3.74\times10^{-3}$ $[3.68,3.77]\times10^{-3}$
& $0.898$ $[0.897,0.899]$
& $0.0110$ $[0.00966,0.0115]$ \\
\bottomrule
\end{tabular}
\end{table}

The improvement with data is visible in all three metrics. Increasing the training set from $512$ to $8192$ strings lowers the median forward KL by more than an order of magnitude. The conditional MAE drops from $0.108$ to $0.0110$, so the gain is not only a better total log probability; the model learns a much more accurate next-bit rule.

The seed spread is informative. At $M=2048$, one seed performs much worse than the other two. At $M=8192$, all three seeds give nearly the same result. More data therefore improves both accuracy and optimization stability for this fixed architecture.

For the representative $M=8192$ model, $8192$ fresh ancestral samples give
\begin{equation}
 d_{\mathrm{KL}}^{\leftarrow}=7.65\times10^{-3}\ \text{bits/site},
 \label{eq:reverse-kl-result}
\end{equation}
with a magnetization RMSE of $1.47\times10^{-2}$ and a connected-$ZZ$ correlation RMSE of $1.57\times10^{-2}$. These checks use samples generated by the GRU rather than target strings passed through the network. The learned model therefore also works as a sampler.

A hidden vector of dimension $16$ is therefore enough to learn this long-memory distribution at $N=64$ from finite data. This is a statement about probabilities only. The GRU learns $|\psi(x)|^2$, not the relative phases of $\psi(x)$. The Bell-basis test in \cref{sec:bell-phase} gives a direct example of information that this training objective cannot recover.

\subsection{Generic variational recurrent state}
\label{sec:generic-rnn-results}

\subsubsection{Setup}

The third experiment asks whether a recurrent neural state can learn a complex ground state by VMC. At $N=64$ and $\theta=0.35$, this model is a tanh RNN: each site takes its spin and normalized position and adds two real outputs to $\log\psi$. Runs use local Monte Carlo with $1024$ samples per update, $16$ chains, Adam, no stochastic reconfiguration; hidden widths are $64$ and $128$.

These runs were performed as development tests with a limited optimization time. The recorded runs had not converged at their final points, and the energy was still decreasing. I therefore use them only to compare optimization behavior and do not treat the final energies as converged results.

\subsubsection{Changes made between runs}

The first hidden-$64$ run used learning rate $5\times10^{-3}$. It descended quickly and reached the best recorded error, but the job ended after the $200$-update checkpoint. I then lowered the learning rate to $10^{-3}$ to test whether a more stable step would help. The descent was smoother but much slower: after $300$ updates the error was still larger than in the higher-learning-rate run.

A hidden width of $128$ with learning rate $3\times10^{-3}$ was tested to see whether more recurrent memory would solve the problem. It did not give a clear short-budget improvement and made each update more expensive. I therefore returned to hidden width $64$ and used $3\times10^{-3}$ as a compromise between the two earlier learning rates. That run also kept improving through its last checkpoint.

This sequence changed the optimizer scale and hidden width, but it did not change the basic difficulty: the network had to learn a useful hidden coordinate, a recurrent update rule, and the complex phase structure from noisy sampled gradients. The MPS-induced model changes the architecture itself instead of only tuning these hyperparameters.

\begin{table}[!htbp]
\centering
\small
\caption{Last retained checkpoint from each generic recurrent VMC run. The Monte Carlo uncertainty $\sigma$ is the uncertainty of the sampled energy estimate at that checkpoint.}
\label{tab:generic-rnn-checkpoints}
\begin{tabular}{@{}ccccc@{}}
\toprule
Hidden width & learning rate & update & $E-E_0$ & $\sigma$ \\
\midrule
$64$  & $5\times10^{-3}$ & $200$ & $0.5654$ & $0.0447$ \\
$64$  & $10^{-3}$        & $300$ & $0.9718$ & $0.0628$ \\
$128$ & $3\times10^{-3}$ & $100$ & $1.5545$ & $0.0633$ \\
$64$  & $3\times10^{-3}$ & $300$ & $0.6118$ & $0.0422$ \\
\bottomrule
\end{tabular}
\end{table}

The best recorded value is
\begin{equation}
 \Delta E\simeq0.5654.
 \label{eq:generic-rnn-best}
\end{equation}

The energy keeps falling, but much more slowly than for the structure-matched model in \cref{fig:structured-convergence}. Longer runs with this fixed architecture and sampler, multiple seeds, and a proper stopping rule would be needed to determine the final error floor. The comparison with \cref{fig:structured-convergence} also mixes architecture with gradient source: the generic runs use sampled VMC gradients, whereas the MPS-induced model is optimized with exact contractions. These runs therefore indicate a trend but do not by themselves isolate the effect of the architecture.

\subsection{Comparison of the three neural experiments}

\begin{table}[H]
\centering
\small
\caption{The neural calculations answer different questions.}
\label{tab:neural-summary}
\begin{tabularx}{\textwidth}{@{}P{0.20\textwidth}P{0.21\textwidth}P{0.21\textwidth}Z@{}}
\toprule
Method & Training signal & Learned object & Main result \\
\midrule
Full-basis RBM & exact energy at $N=8$ & complex wavefunction & median energy-density error $6.03\times10^{-8}$ \\
Data-driven GRU & $N=64$ measurement strings & Born distribution & at $M=8192$, median KL $3.74\times10^{-3}$ bits/site and $F_{\mathrm{cl}}=0.898$ \\
Generic recurrent VMC & sampled local energies & complex wavefunction & best recorded $E-E_0\simeq0.565$; still decreasing when the runs stop \\
\bottomrule
\end{tabularx}
\end{table}

The RBM reaches the small state under an exact objective, and the GRU learns the $N=64$ measured distribution from a few thousand strings. The generic variational runs improve slowly, but since they did not converge, they do not show that this ansatz fails.

%% file: sections/05_geometry_and_implications.tex
\section{Why the CMI changes with the rotation angle}
\label{sec:angle-interpretation}

This section gives a simpler physical picture. Each measurement outcome acts on the two-dimensional MPS bond. The rotation angle changes whether that action preserves, filters, or erases the incoming virtual information. The final CMI also depends on whether the endpoint measurement can read the information that survives.

\subsection{One measurement on the virtual bond}
\label{sec:one-bond-map}

For one representative bulk outcome in the HES frame, the conditioned virtual map is
\begin{equation}
 B^{\mathrm{ES}}_0(\theta)
 =\frac{1}{\sqrt2}
 \left[\cos\!\left(\frac{\theta}{2}\right)\id
      +\sin\!\left(\frac{\theta}{2}\right)X\right].
 \label{eq:hes-outcome-map}
\end{equation}
The neighboring sublattice has the same form with $Z$ replacing $X$. A singular-value decomposition writes any matrix as $B=U\,\mathrm{diag}(\sigma_1,\sigma_2)V^{\dagger}$, where $U$ and $V$ are unitary and $\sigma_1\ge\sigma_2\ge0$. The singular values say how strongly two orthogonal virtual directions are transmitted.

For \cref{eq:hes-outcome-map},
\begin{equation}
 \sigma_{\pm}^{\mathrm{ES}}(\theta)
 =\frac{1}{\sqrt2}
 \left|\cos\!\left(\frac{\theta}{2}\right)
 \pm\sin\!\left(\frac{\theta}{2}\right)\right|.
 \label{eq:hes-singular-values}
\end{equation}
At $\theta=0$, the two singular values are equal. The map is proportional to a unitary and preserves both virtual directions equally. At $\theta=\pi/2$, the smaller singular value is zero. The map has rank one, so different incoming bond states are compressed onto the same outgoing direction.

A useful dimensionless measure of this filtering is
\begin{equation}
 a(B)=\frac{\sigma_2(B)}{\sigma_1(B)},
 \qquad 0\le a(B)\le1.
 \label{eq:anisotropy-ratio}
\end{equation}
The value $a=1$ means no directional filtering, while $a=0$ means complete erasure of one direction. The determinant gives the product of the singular values. For the map above,
\begin{equation}
 \left|\det B^{\mathrm{ES}}_0(\theta)\right|
 =\frac12|\cos\theta|.
 \label{eq:hes-determinant}
\end{equation}
The same factor $|\cos\theta|$ appears in the endpoint upper bound in \cref{eq:yang-bound}, although a one-step determinant alone is not the full CMI.

\begin{figure}[!htbp]
\centering
\includegraphics[width=0.78\textwidth]{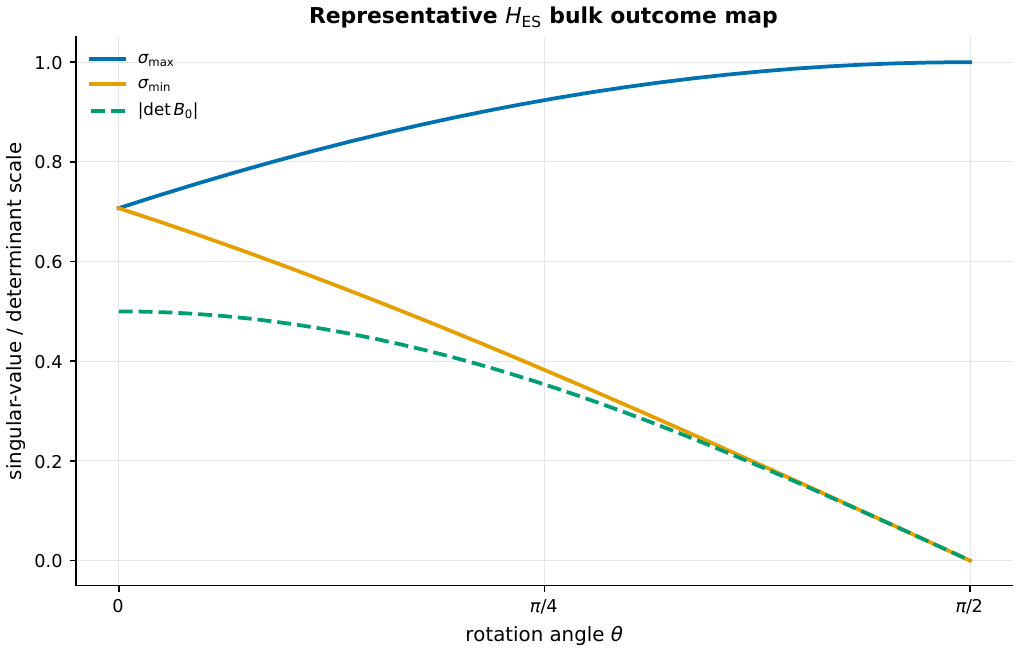}
\caption{Singular values and determinant magnitude for one HES outcome map. Equal singular values at $\theta=0$ mean that both virtual directions survive equally. At $\theta=\pi/2$, one singular value becomes zero and the map has rank one.}
\label{fig:hes-singular-values}
\end{figure}

The conventional cluster frame has the opposite endpoint behaviour: its local computational-basis maps are rank one at $\theta=0$ and proportional to unitaries at $\theta=\pi/2$. The Hadamards relating the two Hamiltonians exchange which physical basis preserves the virtual information and which basis resets it.

This local statement is not yet the CMI. In the conventional frame, the stationary CMI is zero at $\theta=\pi/2$ even though the local maps are reversible. The bond can carry information that the chosen endpoint observable does not distinguish.

\subsection{The whole measurement record}
\label{sec:whole-record}

For a buffer record $w=x_1x_2\cdots x_m$, the virtual bond experiences the ordered product
\begin{equation}
 W_w=B_{x_m}B_{x_{m-1}}\cdots B_{x_1}.
 \label{eq:record-product}
\end{equation}
The order matters because outcome maps generally do not commute. A sequence of weak filters can amplify one direction, rotate it, or erase it. The effect of a long record cannot be read from one local singular value alone.

After the buffer is measured, two separate questions remain.

The first is memory survival: do two different left-endpoint conditions still produce distinguishable virtual states after $W_w$? The second is readout: does the right-endpoint measurement distinguish those surviving states? CMI includes both questions because
\begin{equation}
 I_m=\sum_w p(w) I(A:C\mid W=w).
 \label{eq:cmi-record-average}
\end{equation}
Records that preserve information matter only in proportion to their probability, and preserved virtual information contributes only if the endpoint measurement can see it.

The full calculation can be summarized as
\begin{equation}
 \theta
 \longrightarrow \{B_x(\theta)\}
 \longrightarrow W_w(\theta)
 \longrightarrow p(c\mid a,w)
 \longrightarrow I_m.
 \label{eq:cmi-calculation-chain}
\end{equation}
The bond dimension stays two throughout this chain. What changes with $\theta$ is the lifetime and visibility of the information inside that two-dimensional space.

\subsection{A Bell measurement can see phase information}
\label{sec:bell-phase}

The data-driven GRU learns the computational-basis probabilities $|\psi(x)|^2$. It can match those probabilities while missing signs or complex phases. I test this directly by changing the measurement basis on pairs of sites.

For $N=8$, I measure sites $1$ and $8$ in the $Z$ basis and measure the three middle pairs $(2,3)$, $(4,5)$, and $(6,7)$ in the Bell basis. For a Bell outcome $\beta$, the induced two-site virtual map is
\begin{equation}
 M_{\beta}
 =\sum_{x,y\in\{0,1\}}
 \braket{\beta|x,y}\,A_j^x A_{j+1}^y.
 \label{eq:bell-virtual-map}
\end{equation}
Unlike a computational-basis outcome, this is a coherent sum of several tensor products. Relative signs therefore affect the probability.

As a control, I replace the original amplitudes by
\begin{equation}
 \psi_{\mathrm{abs}}(x)=|\psi(x)|.
 \label{eq:sign-erased-state}
\end{equation}
The state remains normalized and has exactly the same computational-basis distribution as the original state, but its sign pattern is erased.

The four Bell states are
\begin{equation}
 \ket{\Phi^{\pm}}=\frac{\ket{00}\pm\ket{11}}{\sqrt2},
 \qquad
 \ket{\Psi^{\pm}}=\frac{\ket{01}\pm\ket{10}}{\sqrt2}.
 \label{eq:bell-states}
\end{equation}
Fix the outcomes $r$ on all other sites. The probability of a $\Phi^{\pm}$ outcome is
\begin{align}
 p(\Phi^{\pm},r)
 &=\frac12\left|\psi_{00,r}\pm\psi_{11,r}\right|^2
 \label{eq:bell-prob-compact}\\
 &=\frac12\left(
 |\psi_{00,r}|^2+|\psi_{11,r}|^2
 \pm2\operatorname{Re}[\psi_{00,r}\psi_{11,r}^{*}]
 \right).
 \label{eq:bell-prob-expanded}
\end{align}
The same expression holds for $\Psi^{\pm}$ with $00,11$ replaced by $01,10$. The first two terms in \cref{eq:bell-prob-expanded} are fixed by computational-basis probabilities. The last term is an interference term and depends on relative phase.

\begin{figure}[!htbp]
\centering
\includegraphics[width=0.84\textwidth]{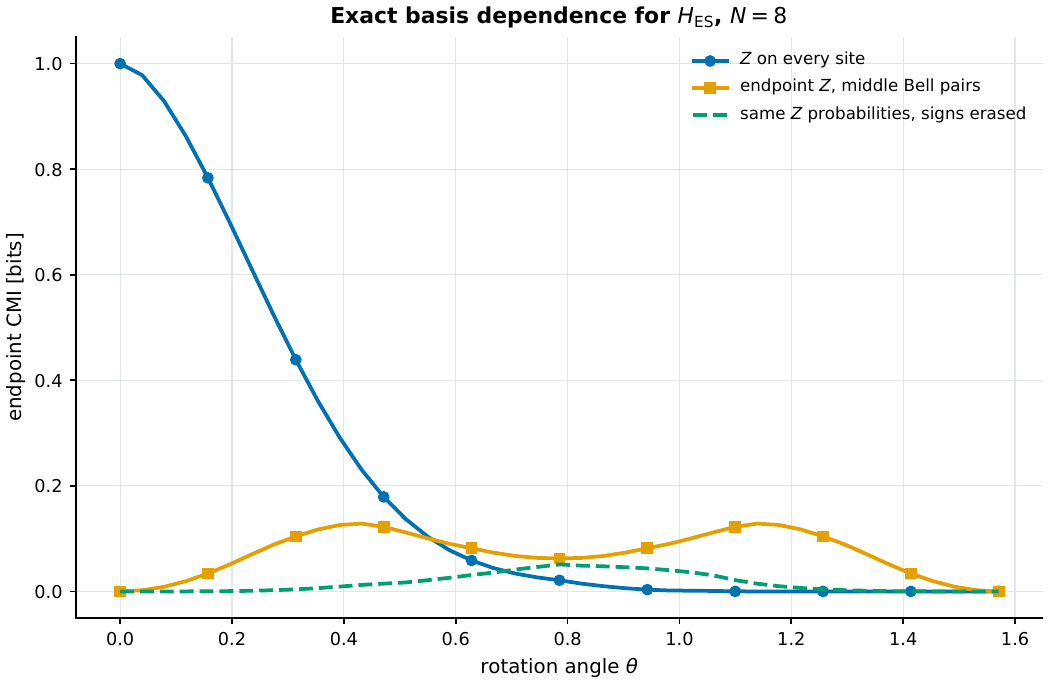}
\caption{Exact $N=8$ basis dependence for the rotated HES state. The all-$Z$ endpoint CMI decreases from one bit to zero. Bell measurements on the middle pairs give an interior signal. Replacing each amplitude by its magnitude leaves the $Z$-basis distribution unchanged but changes the Bell-readout CMI.}
\label{fig:mixed-basis-cmi}
\end{figure}

In density-matrix language, a computational-basis measurement sees diagonal entries, while the Bell measurement also probes off-diagonal coherence. The difference between the original and sign-erased curves in \cref{fig:mixed-basis-cmi} is therefore more useful than the exact location of a small finite-size peak. It proves that the same $Z$-basis data can correspond to different coherent readout statistics.

The Bell-readout curve is not required to decrease monotonically with $\theta$. The rotation first mixes the cluster-state sign structure into combinations that a Bell measurement can read. At the same time, increasing $\theta$ makes the HES virtual maps more anisotropic and finally rank one. The signal is controlled by a competition between phase accessibility and memory survival. The finite-$N$ peak positions are details of this chosen protocol, not special universal angles.

For the learning problem, good likelihood in one basis does not imply a correct wavefunction. A probability model can reproduce all diagonal observables in that basis and still fail for a coherent measurement. A complex variational state must learn both the Born probabilities and the relative phases.

\subsection{Memory size and memory lifetime}
\label{sec:three-memory-notions}

The cluster results separate memory size from memory lifetime. The exact MPS has bond dimension
\begin{equation}
D=2
\end{equation}
for all chain lengths and rotation angles considered here. In contrast, the CMI sequence $I_m$ can remain nonzero over a long range of buffer lengths. Thus, a long CMI lifetime does not require a large virtual state: the same two-dimensional virtual space can carry information over many sites.

\subsection{Connection with measurement-based quantum computation}

The virtual degrees of freedom in the MPS have a direct interpretation in measurement-based quantum computation (MBQC). For a physical measurement outcome $s$, the corresponding MPS tensor acts on the virtual space as
\begin{equation}
    |v_{j+1}\rangle
    \propto
    A^{s_j}|v_j\rangle .
\end{equation}
Thus, a sequence of physical measurements generates a sequence of outcome-dependent maps
\begin{equation}
    |v_m\rangle
    \propto
    A^{s_m}\cdots A^{s_2}A^{s_1}|v_0\rangle .
\end{equation}
In the cluster wire, the virtual space has dimension $D=2$, so it can be viewed as a logical qubit. The measurement record determines how this logical state is propagated and which byproduct operation has to be tracked \cite{GrossEisertSchuchPerezGarcia2007,Wei2021MBQC}.

For a measurement outcome $s$, if
\begin{equation}
    A^s = c_s U_s ,
\end{equation}
with $U_s$ unitary, then the measurement implements a unitary logical operation up to the scalar factor $c_s$. The scalar affects the probability of obtaining outcome $s$, while $U_s$ determines the logical update. The outcome-dependent operation can therefore be corrected or tracked as a known byproduct. At the unrotated cluster-wire point, this update can be reduced to a Pauli-frame variable. For example, a sequence of binary outcomes can be summarized by the parity
\begin{equation}
    p_m = s_1\oplus s_2\oplus\cdots\oplus s_m .
\end{equation}
Although the full measurement record contains $m$ bits, the logical frame only requires this one-bit variable.

The opposite case is a rank-deficient measurement map. In particular, for
\begin{equation}
    A^s = |u_s\rangle\langle w_s|,
\end{equation}
we have
\begin{equation}
    A^s|v\rangle
    =
    |u_s\rangle\langle w_s|v\rangle .
\end{equation}
After normalization, the outgoing virtual state is independent of the incoming direction. Such an outcome therefore removes the corresponding logical degree of freedom rather than propagating it. This gives a useful interpretation of the rank of the MPS measurement map: full-rank maps can retain the logical degree of freedom, whereas a rank-one map acts as a reset of that degree of freedom.

This MBQC interpretation also clarifies what is and is not measured by the CMI in this work. The virtual state $|v_m\rangle$ describes the quantum information propagated through the MPS, whereas
\begin{equation}
    I(S_0:S_{m+1}\mid S_1,\ldots,S_m)
\end{equation}
is a property of the classical measurement record. The two quantities are related because the measurement outcomes determine the virtual updates, but they are not identical. In particular, a logical degree of freedom can remain in the virtual bond while a chosen endpoint measurement basis fails to distinguish it. Therefore, reversibility of the virtual update does not by itself imply a nonzero endpoint CMI.

This distinction is also important for the relation to symmetry-protected topological (SPT) order. The cluster and AKLT states belong to nontrivial SPT phases \cite{Pollmann2012,ElseBartlettDoherty2012}, but the CMI studied here is not an SPT invariant. It depends on the measurement basis, the boundary conditions, and the choice of endpoint variables. A change in these measurement settings can therefore change the observed CMI without changing the underlying quantum phase.

%% file: sections/07_discussion_conclusion.tex
\section{Discussion and outlook}
\label{sec:discussion}

\subsection{Summary of results}
\label{sec:findings}

The rotated cluster chain has bond dimension $D=2$ at every length and every angle, yet its entanglement-swapping readout carries one bit of endpoint information across an arbitrarily long buffer at $\theta=0$, and this endpoint CMI falls to zero at $\theta=\pi/2$. The stationary bulk readout of the locally equivalent conventional cluster state behaves differently: it vanishes at both endpoint angles and is nonzero only in between. For the AKLT chain the CMI is exactly $I_m=\tfrac{4}{9}\,3^{-m}$, and its decay length coincides with the two-point correlation length because the probability of the only memory-preserving outcome equals the subleading transfer eigenvalue. In all three cases the CMI length depends on the measurement basis, the boundary condition, and the partition, and not on the state alone.

On the learning side, the complex RBM reaches the $N=8$ ground state to an energy-density error of order $10^{-8}$ when the energy and its gradient are computed exactly. A GRU with a sixteen-dimensional hidden state learns the $N=64$ computational-basis distribution from a few thousand strings, although, as the Bell readout shows, these data do not fix the phases. The generic recurrent ansatz trained by VMC was still improving when its runs ended. The MPS-induced model with $\chi=2$, which has $48$ real parameters, reaches the ground state to near machine precision, while the $\chi=1$ control does not.

\subsection{Discussion}
\label{sec:significance}

Measurement records from quantum simulators often show long-range conditional structure, and it is natural to read such structure as a sign of a large or highly entangled state. The cluster chain is a simple counterexample. A bond of dimension two is enough to carry conditional information over any distance, and because the outcome-averaged transfer channel is unchanged by the local rotation (\cref{sec:transfer-channel}), the ordinary correlations stay short while the CMI changes strongly with angle. A long CMI therefore indicates that a small amount of information is protected and that the chosen readout can see it. It does not by itself indicate a large bond dimension.

For neural quantum states, the useful feature of this family is that its exact bond dimension is known. Any difficulty a model has in reaching the ground state must then come from optimization, sampling, or parameterization, and not from a lack of expressive power. The present comparison between the generic recurrent ansatz and the MPS-induced model points to a large effect of building the correct virtual memory into the architecture. It is not yet a controlled measurement of that effect, since the two models were also trained with different gradients, one sampled and one exact. The Bell-basis calculation addresses a separate question. Because the sign-erased state has the same computational-basis distribution as the original but a different Bell-readout CMI, it gives a small, exactly computable test that any model trained only on computational-basis data must fail.

\subsection{Future directions}
\label{sec:outlook}

The most direct extension is a controlled version of the generic recurrent benchmark. A single architecture, sampler, initialization, optimizer, and stopping rule should be run over several seeds at each angle until the energy reaches a plateau. Besides the energy, such runs should record the energy variance
\begin{equation}
\operatorname{Var}(H)=\langle H^2\rangle-\langle H\rangle^2,
\label{eq:varH}
\end{equation}
the Monte Carlo autocorrelation time, and the effective sample size. Training the $\chi=2$ MPS-induced model with the same sampler would separate the effect of the architecture from that of gradient noise. Adding stochastic reconfiguration, or replacing MCMC by an autoregressive complex ansatz with exact sampling, would further isolate the role of the optimizer.

For the data-driven task, the next step is to train on strings from several measurement bases and test on a basis not used in training. The Bell readout is a natural test target because its CMI is exactly computable here. This experiment would show how many bases and samples are needed to fix the phases, and would connect the benchmark to the neural quantum state tomography literature.

The structure-matched model performs well here because its built-in bond dimension is exact. Perturbing the cluster Hamiltonian, for example with a transverse field, or moving to MPS families with larger bond dimension, would show how this advantage changes once the required bond dimension exceeds the built-in one. The same calculations would show how the conditional structure of the measured process changes away from the stabilizer point.

The results also raise the question of which local measurement basis maximizes the CMI lifetime of a given MPS, and how such optimal readouts are related to string order and to the symmetries that protect the cluster and AKLT phases. Finally, realistic measurement records are noisy and finite. Decoherence makes the state mixed, so the pure-state identity of \cref{sec:cmi-lengths} no longer forces $I(A{:}C|B)=I(A{:}C)$, and small CMI values are hard to estimate from finite samples because entropy estimators are biased where the signal is smallest. Both issues would need to be addressed before CMI can be used routinely on experimental data. Two-dimensional cluster states, where exact contraction is no longer available, are a further natural test.

\subsection{Conclusion}
\label{sec:finalconclusion}

The rotated cluster chain shows that a state with a two-dimensional bond can produce measurement records with long conditional memory, and that whether this memory is visible depends on the readout. For learning, the main difficulty observed here was not the size of the required memory but finding a hidden-state update with the right structure: a model that has this structure built in reaches the ground state easily, while a generic recurrent model trained by VMC improves slowly. Exact MPS calculations make it possible to check these statements quantitatively, and the same approach can be applied to other states with known tensor-network representations.

%% file: sections/06_methods_and_limitations.tex
\section{Numerical checks and limitations}
\label{sec:checks}
 
This appendix lists the checks behind the numbers quoted in the main text
and the limitations of those numbers.
 
\subsection{Validation}
\label{sec:validation}
 
All cluster and AKLT probabilities come from explicit MPS tensors, contracted
from the physical boundaries for finite chains and from transfer-channel fixed
points for stationary bulk blocks. Every joint table is checked for
normalization, $\delta_{\mathrm{norm}}=\left|\sum_x p(x)-1\right|$, and
non-negativity before an entropy is computed; only roundoff-level negative
entries are clipped. The sequential sampler is verified against the
conditional-sampling identity of \cref{sec:mps-sampling}: the product of conditional
probabilities must reproduce the direct contraction of the completed string.
A CMI length is fitted only to an exact sequence $I_m$, with even and odd
buffers inspected separately, and no length is quoted when the signal is
already at numerical precision; the exponential bound of \cref{sec:cmi-scan} is
never substituted for measured CMI.
 
The reference states pass independent cross-checks. The explicit HES MPS
agrees with sparse exact diagonalization at $N=4,6,8$ with absolute overlaps
of at least $0.9999999999999997$, energies equal to $-N$, and a largest
stabilizer residual of $1.11\times10^{-16}$. Because the Hamiltonian is
frustration free, I also monitor $\delta_K=\max_j|1-\langle K_j\rangle|$,
which tests every term separately instead of allowing cancellations inside
the total energy; the trained $\chi=2$ model keeps all stabilizer
expectations above $0.999999999956$. The CMI implementation reproduces three
exact limits: the one-bit HES endpoint value at $\theta=0$, the vanishing
endpoint value at $\theta=\pi/2$, and the AKLT law of \cref{sec:aklt-control}. The
data-driven GRU is checked in three ways: held-out likelihood,
exact target conditional probabilities, and observables from freshly
generated samples.
\subsection{Training protocols}
\label{sec:protocols}
 
\begin{table}[H]
\centering
\small
\caption{Training protocols used in the numerical experiments.}
\label{tab:protocols}
\begin{tabularx}{\textwidth}{@{}P{0.20\textwidth}P{0.18\textwidth}P{0.15\textwidth}Z@{}}
\toprule
Model & Gradient source & Repetitions & Optimization and stopping rule \\
\midrule
Complex RBM, $N=8$ & exact full-basis energy & five seeds & 2000 Adam steps, then deterministic L-BFGS refinement \\
Positive GRU, $N=64$ & mini-batch likelihood & three seeds per data size & AdamW, held-out validation, early stopping, maximum 35 epochs \\
Generic recurrent state, $N=64$ & sampled local energies with local MCMC & several development runs & Adam; jobs stopped by a 30-minute queue limit rather than convergence \\
MPS-induced recurrent model, $N=64$ & exact transfer-contracted energy & five $\chi=2$ seeds plus one $\chi=1$ control & 150 Adam steps, tensor-norm projection, then L-BFGS refinement \\
\bottomrule
\end{tabularx}
\end{table}
 
The RBM and MPS-induced models optimize deterministic double-precision
objectives, so their seed spread measures optimization robustness rather than
Monte Carlo noise; Adam locates the basin and L-BFGS refines it. The GRU uses
disjoint training, validation, and test strings, with model selection on
validation likelihood only and no exact tensor inserted into its parameters.
The generic recurrent VMC gradients are sampled and correlated, with no
stochastic reconfiguration, which makes those runs a pilot rather than a
matched comparison with the deterministic MPS optimization.
 
\subsection{Limitations}
\label{sec:limitations}
 
The $\chi=2$ recurrent MPS
is matched to a state whose bond dimension is known to be two; it
shows the value of a correct inductive bias and says nothing about MPS ans\"atze
in general, and its exact-gradient optimization is not matched to the sampled gradients of the generic VMC runs. The RBM benchmark is exact but limited to $N=8$ full-basis summation
and says nothing about sampled large-$N$ scaling. The generic VMC runs are
truncated by the wall-time limit and do not form a controlled seed sweep, so
they establish no error floor; they only show slow, noisy improvement. The
GRU study uses one small architecture, three seeds, noiseless synthetic data,
at most $8192$ strings, and a single basis and angle; it does not recover
phases, as the $N=8$ Bell test makes explicit. The CMI scans reach $m\le12$
(stationary) and $m\le14$ (endpoint), so the quoted lengths are finite-window
summaries that cannot resolve the asymptotic tail near a vanishing signal.
Finally, the two cluster CMI curves use different partitions (stationary
bulk and physical endpoints) and are different observables, so their
difference is not a disagreement between numerical methods.